\documentclass[11pt, a4paper]{article}

\usepackage[utf8]{inputenc}
\usepackage[T1]{fontenc}
\usepackage{amsmath, amssymb}
\usepackage{graphicx}
\usepackage[margin=2.5cm]{geometry}
\usepackage{authblk} 
\usepackage{setspace} 
\usepackage{lineno}   
\usepackage{hyperref} 
\usepackage{xcolor}   
\usepackage{booktabs}
\usepackage{tabularx}
\usepackage{caption}
\usepackage{doi}

\hypersetup{
    colorlinks=true,
    linkcolor=blue,
    filecolor=magenta,      
    urlcolor=cyan,
    citecolor=blue,
}
\title{\textbf{Multi-tesla operation of high-temperature superconducting cavities for accelerated axion dark matter searches}}

\author[1,a]{Danho Ahn}
\author[1,b]{Jinsu Kim}
\author[1]{Seongtae Park}
\author[2,1,b]{Jiwon Lee}
\author[1,b,*]{Ohjoon Kwon}
\author[1,*]{Woohyun Chung}
\author[1,d]{HeeSu Byun}
\author[1,b]{Sergey Uchaikin}
\author[3]{Arjan Ferdinand van Loo}
\author[3,4]{Yasunobu Nakamura}
\author[2]{Dojun Youm}
\author[1,b]{SungWoo Youn}
\author[1,2,c]{Yannis K. Semertzidis}

\affil[1]{Center for Axion and Precision Physics Research, Institute for Basic Science\\
Daejeon 34051, Republic of Korea}
\affil[2]{Department of Physics, Korea Advanced Institute of Science and Technology (KAIST)\\
Daejeon 34141, Republic of Korea}
\affil[3]{RIKEN Center for Quantum Computing (RQC), Wako, Saitama 351-0198, Japan}
\affil[4]{Department of Applied Physics, Graduate School of Engineering \\
The University of Tokyo, Bunkyo-ku, Tokyo 113-8656, Japan}

\affil[a]{\textit{Present address:} INFN-Sezione di Padova, Via Marzolo 8, 35131, Padova, Italy}
\affil[b]{\textit{Present address:} Dark Matter Axion Group, Institute for Basic Science, Daejeon 34051, Republic of Korea}
\affil[c]{\textit{Present address:} Innovative Solutions R\&D, LLC, Stony Brook, NY 11790, USA}
\affil[d]{\textit{Present address:} Max-Planck-Institut für Physik, Garching, Germany}

\affil[*]{These authors contributed equally to this work as corresponding authors. e-mail: o1tough@ibs.re.kr; gnuhcw@gmail.com}

\date{\today}

\let\combinedsavedtitle\title
\let\combinedsavedauthor\author
\let\combinedsaveddate\date

\begin{document}

\maketitle

\begin{abstract}

Axion haloscopes search for dark matter through its resonant conversion to microwave photons in a strong magnetic field. Improving cavity quality factors under multi-tesla fields can increase their sensitivity, but vortex losses and assembly interfaces constrain superconducting implementations. Here we develop three-dimensional cavities from mechanically delaminated rare-earth barium copper oxide (REBCO) tapes, removing the metallic substrate while retaining the copper stabilizer as a conducting support. Longitudinal segmentation aligns the seams with the dominant surface-current paths, while the copper backing increases the conducting depth at the gaps. Employing a two-track strategy, we first unveiled the intrinsic high-field potential of the material with a $5.4\,\text{GHz}$ resonant cavity, achieving an unloaded quality factor $Q_0$ of $1.4 \times 10^7$ in an $8\,\text{T}$ magnetic field---exceeding conventional copper baselines by two orders of magnitude. Second, prioritizing practical haloscope integration, we engineered a frequency-tunable, volume-maximized $2.3\,\text{GHz}$ pathfinder cavity. Deployed in the Pilot Axion Cavity Experiment at the Center for Axion and Precision Physics Research (CAPP-PACE), this system achieved a $180\,\text{mK}$ noise temperature and a fivefold loaded-quality factor enhancement over the copper reference, corresponding to an estimated 8.3-fold improvement in the analysis-adjusted scan-rate figure of merit. The search established the most stringent axion--photon coupling limits to date around $9.46\,\mu\mathrm{eV}$ at 90\% confidence level.

\end{abstract}

\section*{Introduction}
The nature of dark matter and the strong charge-parity ($CP$) problem in quantum-chromodynamics (QCD) remain central mysteries in fundamental physics~\cite{Rubin1970Andromeda, tHooft1976symmetry, abel2020measurement, kim2010axions}. The hypothetical axion---a pseudo-Nambu-Goldstone boson predicted by Peccei-Quinn symmetry~\cite{peccei1977cp, weinberg1978new, wilczek1978problem}---offers a compelling unified solution to both puzzles, providing a viable cold dark matter candidate produced via early-universe vacuum misalignment~\cite{Preskill1983Axion, Abbott1983Axion, Dine1983Axion, Marsh2016AxionCosmology}. To search for these particles, cavity haloscopes exploit Sikivie's resonant conversion mechanism, converting axion dark matter into microwave photons within a resonant cavity immersed in a strong static magnetic field~\cite{sikivie1983experimental}.

The scan rate of a haloscope follows from its signal power and noise through the radiometer equation~\cite{dicke1946}. For conventional cavities, the scan rate typically scales as $d\nu/dt \propto V^2 B^4 Q_L T_{\text{sys}}^{-2}$, where $V$ is the effective cavity volume, $B$ is the magnetic field, $Q$ is the cavity quality factor, and $T_{\text{sys}}$ is the system noise temperature~\cite{Kim2020AxionScanRate}. High-field magnets and quantum-noise-limited amplifiers have improved experimental sensitivity~\cite{BackesK2021Haystacaxion, yi2023axion, Kim2023NearQuantum, ahn2024prx}. Cryogenic copper losses, including anomalous-skin-effect behavior, nevertheless limit cavity performance in a geometry- and frequency-dependent manner~\cite{anormalous1950nat}. Superconducting alternatives must control field-induced vortex dissipation~\cite{Gittleman1966VortexModel}.

High-temperature superconductors (HTS)---specifically biaxially-textured rare-earth barium copper oxide (REBCO) coated conductors---offer a promising alternative owing to their high upper critical fields and strong vortex pinning (e.g., via BaHfO$_3$ artificial pinning centers)~\cite{Romanov2020HighFreq}. 
Their biaxially textured films provide well-aligned paths for low-loss microwave currents, but producing comparable films directly on the curved inner surface of a three-dimensional cavity is challenging. 
We therefore constructed our first polygonal cavity from commercially developed, biaxially textured YBCO tapes, arranging the segments so that the $TM_{010}$ surface currents ran along the tapes rather than across their interfaces~\cite{ahn2022biaxially}. At $6.93\,\text{GHz}$, the cavity maintained an unloaded quality factor approximately six times that of an otherwise identical copper cavity in fields up to $8\,\text{T}$. Recent efforts subsequently demonstrated an HTS-lined cavity for an axion search at $8.84\,\text{GHz}$ and $11.7\,\text{T}$, reporting an approximately $1.5$-fold quality-factor improvement over its copper reference~\cite{RADES2025hts}.

These demonstrations establish the value of HTS tapes in high-field cavities, while leaving an assembly constraint in our segmented $TM_{010}$ design.
The micron-scale superconducting films provide little conducting depth at the segment interfaces. Assembly offsets can allow microwave fields entering the seams to encounter the buffer stack and metallic substrate, adding to the cavity loss. In this work, we address the limitation with segmented cavities assembled from strain-controlled, mechanically delaminated REBCO tapes. By exploiting the low transverse fracture toughness of coated conductors~\cite{vanDerLaan2007Delam, Shin2014Transverse, Yanagisawa2011Cleavage}, we physically strip away the lossy metallic substrate while retaining the thick copper stabilizer ($20\text{--}100\,\mu\text{m}$). The retained copper stabilizer provides a conductive backing within each segment and increases the conducting depth at its edges. This geometry reduces the exposure of non-superconducting tape layers to fields entering the assembly seams.

This conductive-backed platform allows us to navigate the core engineering trade-offs inherent to practical haloscopes. Because the scanning speed scales as $d\nu/dt \propto V^2 Q_L$ assuming other experimental parameters remain constant, pursuing an ultra-high $Q$ is counterproductive if it requires bulky internal support structures that severely sacrifice the cavity volume $V$ or introduce dynamic tuning instabilities. Therefore, we executed a two-pronged strategy. First, to explore the intrinsic physical bounds of the delaminated film, we constructed a fixed-frequency $5.40\,\text{GHz}$ europium barium copper oxide (EuBCO) cavity, breaking the $10^7$ quality factor barrier under multi-tesla fields ($Q_0 = 1.4 \times 10^7$ at $8\,\text{T}$). Second, prioritizing operational haloscope sensitivity, we engineered a tunable, volume-maximized $2.30\,\text{GHz}$ gadolinium barium copper oxide (GdBCO) pathfinder cavity. Integrated into the Pilot Axion Cavity Experiment at the Center for Axion and Precision Physics Research (CAPP-PACE) haloscope equipped with a quantum-noise-limited amplifier~\cite{Kwon2021FirstResults, Kim2023NearQuantum}, this system achieved a near-quantum-limited system noise temperature of $180\,\text{mK}$ and completed a physics run setting exclusion limits over $9.4376{-}9.4914\,\mu\text{eV}$, with the best sensitivity reaching $1.4 \times$ the Kim-Shifman-Vainshtein-Zakharov (KSVZ) benchmark~\cite{kim1979weak,shifman1980CPinv}. By demonstrating a two-order-of-magnitude boost in intrinsic high-field quality factor alongside an operational haloscope whose measured parameters yield an estimated 8.3-fold improvement in the analysis-adjusted scan-rate figure of merit relative to the copper reference, these results firmly establish strain-controlled, conductive-backed HTS cavities as a scalable and practical platform for next-generation dark matter searches.

\section*{Results}

\subsection*{Design and structure of conductive-backed 3D HTS cavities}
To realize these conductive-backed cavities, we exploited the mechanical anisotropy of the multilayer REBCO tape architecture. Despite their high longitudinal load-bearing capacity, REBCO coated conductors are susceptible to interlaminar separation under transverse tensile and peel loading~\cite{vanDerLaan2007Delam, Shin2014Transverse, Yanagisawa2011Cleavage, Lu2025Peel}. The fracture path depends on the conductor architecture and interfacial properties, with delamination reported within the buffer stack, at the buffer/substrate interface, and within the REBCO layer itself~\cite{Hojo2012ModeI, Xin2025TEM}. We used this susceptibility to interlaminar separation as a means of removing the substrate while retaining the REBCO layer on its copper stabilizer.

\begin{figure}[htbp]
    \centering
    \includegraphics[width=0.7\textwidth]{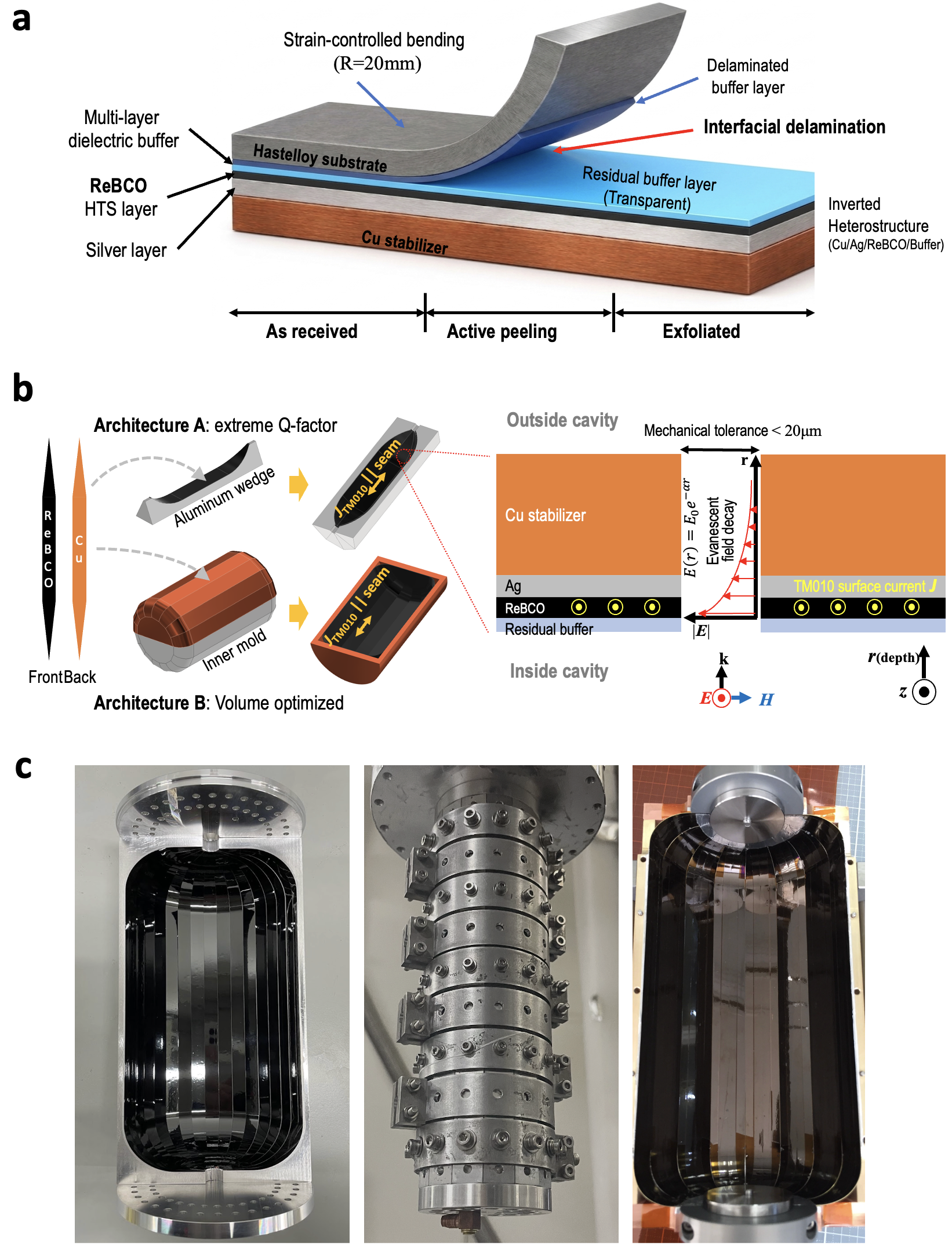}
    \caption{\textbf{Strain-controlled mechanical delamination and waveguide-below-cutoff RF leakage suppression in 3D HTS cavities.} 
    \textbf{(a)} Three-stage mechanical delamination process (\textit{As-received}, \textit{Active peeling}, and \textit{Exfoliated}). Guided peeling around a cylinder ($R = 20\,\text{mm}$) exploits low transverse fracture toughness to cleanly strip the lossy $50\,\mu\text{m}$ Hastelloy substrate along the buffer stack, yielding an inverted heterostructure ($\text{Cu}/\text{Ag}/\text{REBCO}/\text{Buffer}$) backed by a thick copper stabilizer ($20\text{--}100\,\mu\text{m}$). 
    \textbf{(b)} 3D cavity architectures and seam attenuation mechanism. Left: Segmented tape wedges aligned parallel to $TM_{010}$ surface currents ($J_z \parallel \text{seam}$) in precision-wedge (Architecture A) and volume-optimized inner-mold (Architecture B) assemblies. Right: Assembly gap cross-section ($w < 20\,\mu\text{m}$). The Cu-stabilizer depth ($d_{\text{Cu}} \ge 20\,\mu\text{m}$) forms a waveguide below cutoff, exponentially attenuating evanescent RF fields ($E(r) = E_0 e^{-\alpha r}$) along the conductive trench and terminating residual currents on the low-loss cryogenic copper boundary.
    \textbf{(c)} Photographs of fabricated 3D HTS cavities. Left: Partially assembled $2.27\,\text{GHz}$ EuBCO cavity (Architecture A) exposing internal pole-to-pole wedge alignment. Center: Fully assembled $5.40\,\text{GHz}$ EuBCO cavity (Architecture A) outfitted with external mechanical compression clamps to minimize and maintain the seam gap during cryogenic cool-down. Right: Partially assembled $2.30\,\text{GHz}$ GdBCO pathfinder cavity (Architecture B) showing the inner HTS surface and central alignment jig fixtures used to mate the two half-shells.}
    \label{fig:1}
\end{figure}

A controlled mechanical peeling process was applied to remove the lossy $50\,\mu\text{m}$ Hastelloy substrate. During peeling, the tape was guided along a cylindrical jig of radius $R = 20\,\text{mm}$ to control its macroscopic curvature. This yielded an inverted structure in which the REBCO film ($2.5$--$5\,\mu\text{m}$) is supported by the retained copper stabilizer ($d_{\text{Cu}} = 20$--$100\,\mu\text{m}$), as illustrated in Table~\ref{tab:tapes} and Fig.~\ref{fig:1}a. This arrangement places a thick conducting support behind the superconducting surface, allowing the copper thickness to contribute to the conducting sidewall depth at assembly seams, as discussed below. For this application, the relevant performance criterion is the microwave loss of the resulting structure under the cavity operating conditions. High-field microwave measurements showed low surface resistance relative to copper, and a cavity incorporating the exfoliated films achieved $Q_0 = 1.4 \times 10^7$ at $8\,\text{T}$. These results demonstrate that the exfoliated films support low-loss microwave operation in strong magnetic fields.

Qualitative observations also suggested a secondary benefit from residual buffer material on the exposed surface. The peeled surfaces exhibited a reflective appearance across the tape and greater apparent resistance to visible damage during brief chemical exposure than REBCO surfaces exposed by silver removal from tapes without copper stabilization. These observations are consistent with partial retention of the oxide buffer stack and suggest a short-term protective effect during handling and processing. Although the residual material was not characterized quantitatively, the resulting exfoliated structures exhibited low microwave loss under the conditions investigated.

Fundamentally, this inverted architecture resolves the seam dissipation problem via two synergistic mechanisms: topological current alignment and radial evanescent field suppression. Building upon the architectural framework established in our earlier work~\cite{ahn2022biaxially}, the cavities were constructed using pole-to-pole longitudinal wedges (analogous to ``melon slices'') to strictly match the dominant $TM_{010}$ surface current paths ($J_z \parallel \text{seam}$, Fig.~\ref{fig:1}b). This contiguous geometry ensures that the primary assembly seams remain parallel to the surface currents, fundamentally bypassing transverse cross-seam conduction barriers. 

Simultaneously, for the RF electric field ($E_z$) facing the narrow gap openings ($w \approx 10\text{--}20\,\mu\text{m}$), a local idealization of the parallel conducting walls shows that the boundary conditions for this relevant polarization necessitate a transverse variation across the gap. Consequently, the radial path along the gap depth ($r$-direction) acts analogously to a waveguide operating far below its dominant cutoff frequency ($\nu \ll \nu_c \approx c/2w \sim \text{THz}$)~\cite{Jackson1999EM, Pozar2011MW}. This $TM_\text{010}$ electric-field component attempting to leak radially through the gap becomes purely evanescent, decaying exponentially:\begin{equation}E(r) = E_0 \exp(-\alpha r), \quad \text{where } \alpha = \sqrt{\left(\frac{\pi}{w}\right)^2 - \left(\frac{2\pi\nu}{c}\right)^2} \approx \frac{\pi}{w}\end{equation} where $r$ is the radial propagation depth. In standard non-delaminated tapes, the conducting sidewall depth is limited to the superconducting layer ($d_{\text{HTS}} \sim 1.5\,\mu\text{m}$). Because $d_{\text{HTS}} \ll w$, the attenuation factor is negligible. Under such insufficient conducting depth, the evanescent field can couple to external modes and penetrate directly into the highly resistive Hastelloy substrate ($\rho_{\text{Hastelloy}} \approx 125\,\mu\Omega\cdot\text{cm}$), causing severe radiative and ohmic dissipation.

In contrast, our mechanically delaminated architecture extends the conducting sidewall depth to include the thick copper stabilizer ($d_{\text{Cu}} \ge 20\,\mu\text{m}$). This converts the narrow assembly gap into a deep ``conductive trench''. The squared-amplitude attenuation ($A_{\text{gap}}$) of this modeled evanescent component decreases as $\exp(-2\alpha d_{\text{Cu}})$, attenuating macroscopic RF leakage:
\begin{equation} A_{\text{gap}} = \left| \exp\left(-\frac{\pi d_{\text{Cu}}}{w}\right) \right|^2 
\label{eq:gap_attenuation}\end{equation}
For an illustrative conducting depth of $d_{\mathrm{Cu}}=20\,\mu\mathrm{m}$, this model predicts approximately $-55\,\mathrm{dB}$ attenuation for $w=10\,\mu\mathrm{m}$ and $-27\,\mathrm{dB}$ for $w=20\,\mu\mathrm{m}$. These analytical estimates motivate the strict sub-10-$\mu\text{m}$ gap tolerances maintained via precision wedge co-polishing. They serve as an interpretation of the gap mechanics rather than a total computation of cavity leakage, which also depends on finite endcap geometry, junctions, and polarization conversion. While finite-element mesh memory constraints preclude simulating sub-10-$\mu\text{m}$ gaps directly on full 3D models, extrapolations from prior $0.1\,\text{mm}$-gap simulations~\cite{ahn2022biaxially} provide strong architectural context supporting this interpretation. Thus, the continuous copper backing effectively suppresses cross-seam RF leakage, preserving the intrinsic high-$Q$ performance.

To translate this conductive-backed film platform into functional 3D resonant cavities while navigating the practical engineering trade-offs between ultimate quality factor and effective magnet bore volume ($V$), we engineered two distinct architectural variants (Fig.~\ref{fig:1}b, Table~\ref{tab:cavities}).

\textbf{Architecture A: Ultimate quality factor.} 
To push the physical upper bounds of the delaminated films and achieve high quality factors, we assembled EuBCO tapes on precision Aluminum 6061 wedges (Fig.~\ref{fig:1}b). Aluminum was selected for its superior machinability, enabling individual wedges with sub-10-$\mu\text{m}$ tolerances. To overcome the challenge of aligning flexible HTS tapes with sub-10-$\mu\text{m}$ seam precision, delaminated tapes were secured onto individual wedges, after which the joint faces of the tape-wedge assemblies were co-polished as a single unit using fine 2000-grit silicon carbide (SiC) abrasive blocks to guarantee seamless interface mating between adjacent segments. A photograph of a partially assembled $2.27\,\text{GHz}$ EuBCO cavity (Fig.~\ref{fig:1}c, left) displays the internal wedge alignment and pole-to-pole longitudinal seams prior to final enclosure. Because the effective RF penetration depth in the strongly pinned mixed state ($\lambda_{\text{eff}} \sim \mathcal{O}(100)\,\text{nm}$) remains significantly smaller than the $2.5\,\mu\text{m}$ superconducting layer, the RF fields are well screened before reaching the copper boundary, rendering minor thickness variations in the residual stabilizer entirely negligible.

Importantly, our iterative cavity design revealed that the rigidity of the mechanical wedge governs the ultimate limit of seam leakage. In our first-generation $2.27\,\text{GHz}$ EuBCO cavity (No.~3 in Table~\ref{tab:cavities}), the elongated wedge geometry experienced subtle thermal distortion during cryogenic cool-down, yielding an impressive yet seam-bounded $Q_0 \approx 3.5 \times 10^6$ at $8\,\text{T}$ (Fig.~\ref{fig:2}). By scaling to a $5.40\,\text{GHz}$ geometry (No.~4 in Table~\ref{tab:cavities}), the shorter wedge length substantially improved structural stiffness. To prevent thermal gap opening across the seams, external mechanical compression clamps were installed around the cavity body (Fig.~\ref{fig:1}c, center). Although these clamping fixtures slightly increase the external radial footprint (reducing active volume efficiency), they were designed to target sub-$20\,\mu\text{m}$ seam alignment (consistent with standard IT6-IT7 machining grades). This structural containment unlocked an unloaded quality factor of $Q_0 = 1.4 \times 10^7$ at $8\,\text{T}$, establishing a two-order-of-magnitude benchmark over copper (Fig.~\ref{fig:2} and Table~\ref{tab:cavities}).

This performance gain highlights the functional role of the conductive backing compared to non-delaminated designs. In our earlier $6.93\,\text{GHz}$ prototype (No.~1 in Table~\ref{tab:cavities}), a similar segmented wedge architecture constructed from non-delaminated YBCO tapes on Ni-9W substrates was limited to $Q_0 \approx 3.3 \times 10^5$ at $8\,\text{T}$, as RF fields leaked through the narrow assembly gaps into the lossy substrate due to the shallow conducting depth ($d_{\text{HTS}} \sim 0.8\,\mu\text{m} \ll w$), despite adding a separately sputtered silver edge coating to bridge the YBCO surface to the aluminum body~\cite{ahn2022biaxially}. In contrast, utilizing strain-controlled delamination to secure a thick copper stabilizer backing ($d_{\text{Cu}} \ge 20\,\mu\text{m}$) converts the assembly seams into deep waveguides below cutoff. Attenuating this cross-seam leakage below intrinsic film dissipation allows the cavity performance to be governed predominantly by the flux-pinning capability of $c$-axis BaHfO$_3$ nanorods in EuBCO.

\textbf{Architecture B: Volume-Efficient Millikelvin Haloscope.} 
While Architecture A excels in achieving high quality factors, the external fixtures required for aluminum wedges reduce the effective cavity volume within the magnet bore. To maximize active volume while preserving inner RF surface smoothness, we developed an inner-mold fabrication method tailored for haloscope integration (Fig.~\ref{fig:1}b; No.~2 in Table~\ref{tab:cavities}). This monolithic technique required mechanically stiffer tapes to conform to the cylindrical profile; thus, we utilized GdBCO tapes reinforced with a $100\,\mu\text{m}$-thick copper stabilizer layer. 

To execute this assembly, delaminated tapes were precisely aligned over an inner cylindrical mold to form two semi-cylindrical shell halves, with the longitudinal seams secured from the outside using low-temperature indium cold-welding. A photograph of one completed HTS half-shell (Fig.~\ref{fig:1}c, right) displays the smooth inner RF surface and alignment jig fixtures prior to final enclosure. The two half-shells were subsequently mated and mounted into an outer brass/copper frame for structural support and mK thermalization. While this architecture maximizes effective volume ($V = 1.5\,\text{L}$), finite alignment tolerances near the pole convergences inevitably exposed a small fraction ($\sim 5\%$) of the inner RF surface to normal-conducting indium welds and cut edges. Consequently, the operational $Q_0$ of this pathfinder cavity is limited to $\simeq 5\times10^5$ at $8\,\text{T}$ by localized normal-metal dissipation rather than intrinsic seam leakage (Fig.~\ref{fig:2})---delivering a 5-fold improvement over copper while optimizing active volume.

\subsection*{High-field cavity performance in multi-tesla environments}

We characterized the field-dependent surface resistance variation, $\Delta R_s(B) = R_s(B) - R_s(0)$, of delaminated REBCO films using a Hakki-Coleman rutile ($\mathrm{TiO}_2$) dielectric resonator operating near $8\,\text{GHz}$ (see Methods)~\cite{Klein1995Dielectric}. Measurements were performed up to $8\,\text{T}$ in two orthogonal orientations: perpendicular ($B \perp ab$-plane, recorded across $90.0\text{--}208.7\,\text{mK}$) and parallel ($B \parallel ab$-plane, recorded across $85.5\text{--}478.6\,\text{mK}$). During field sweeps, the magnet ramp was paused at selected fields; the absence of systematic $Q$-factor drift during hold intervals confirmed the stability of the extracted field-dependent response on the measurement timescale.

\begin{figure}[htbp]
    \centering
    \includegraphics[width=\textwidth]{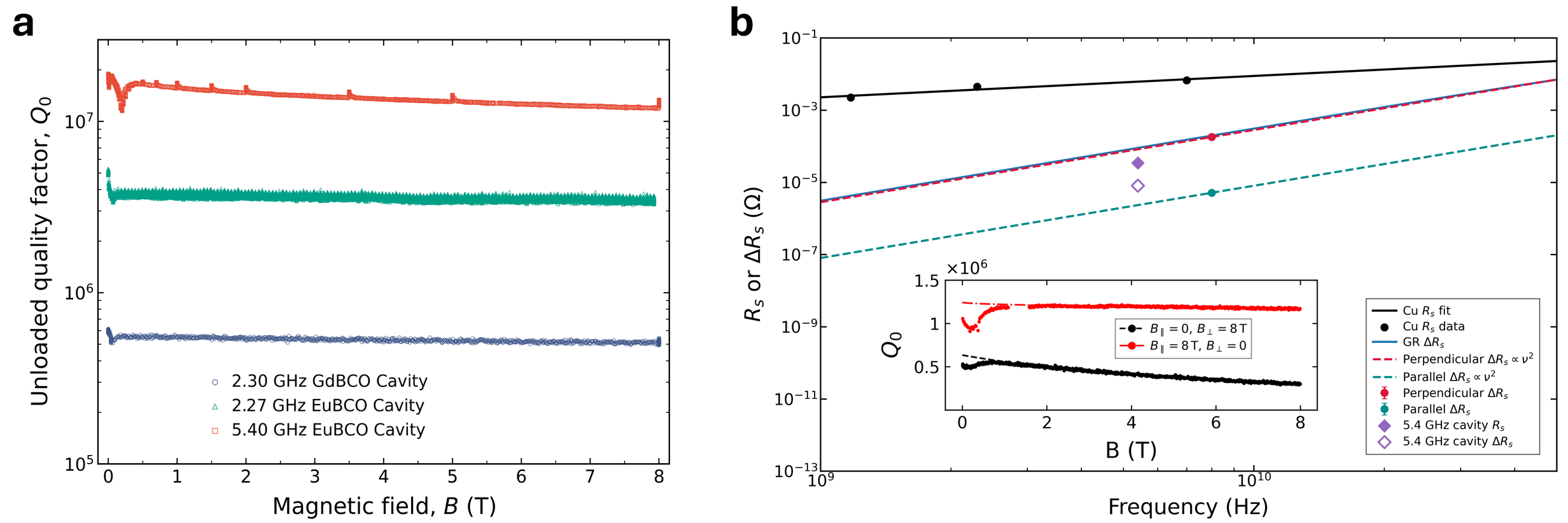}
    \caption{\textbf{High-field quality factor and frequency-dependent surface resistance evaluation.} 
    \textbf{(a)} Measured unloaded quality factor ($Q_0$) as a function of external magnetic field up to $8\,\text{T}$ for all fabricated HTS cavity prototypes: $5.40\,\text{GHz}$ EuBCO (orange, $Q_0 = 1.4 \times 10^7$ at $8\,\text{T}$), $2.27\,\text{GHz}$ EuBCO (teal, $Q_0 \approx 3.5 \times 10^6$ at $8\,\text{T}$), and $2.30\,\text{GHz}$ GdBCO (blue, $Q_0 \approx 5 \times 10^5$ at $8\,\text{T}$; overall axion-run RMS $Q_0 \simeq 4.05 \times 10^5$). 
    \textbf{(b) Main panel:} Frequency dependence of surface resistance ($R_s$) and field-induced resistance increase ($\Delta R_s$) at $B = 8\,\text{T}$ on a log-log scale. Copper $R_s$ data (black circles) and power-law fit ($R_s^{\text{Cu}} \propto \nu^{0.59}$, solid black line) are compared against Gittleman-Rosenblum model (blue solid line) and $\nu^2$ guide lines anchored to the rutile orientation measurements. At $8\,\text{GHz}$, measured $8\,\text{T}$ field-induced resistance increases yield $\Delta R_s^\perp = 179.84 \pm 0.90\,\mu\Omega$ ($B \perp ab$, red circle) and $\Delta R_s^\parallel = 5.13 \pm 0.15\,\mu\Omega$ ($B \parallel ab$, teal circle); errors are formal fit-only uncertainties. For the $5.40\,\text{GHz}$ cavity, the filled purple diamond is $R_{s,\text{eff}}=31.36\,\mu\Omega$, and the open diamond is the adopted background-corrected film value $\Delta R_s=9.06\,\mu\Omega$. 
    \textbf{Inset:} Rutile resonator $Q_0$ versus magnetic field up to $8\,\text{T}$ for Perpendicular ($B \perp ab$) and Parallel ($B \parallel ab$) orientations with nonlinear Gittleman-Rosenblum physical model fits (black dashed and dash-dotted curves) and mapped $\pm 1\sigma$ uncertainty bands. Data points affected by mode crossings in the parallel orientation ($1.087\text{--}1.579\,\text{T}$) were excluded from the analysis.}
    \label{fig:2}
\end{figure}

The microwave performance of Type-II superconductors under magnetic fields is dictated by the dissipative motion of Abrikosov vortices~\cite{Gittleman1966VortexModel}. In the parallel orientation ($B \parallel ab$-plane), intrinsic planar pinning suppresses flux motion, yielding a minimal field-induced increase of only $\Delta R_s^\parallel = 5.13 \pm 0.15\,\mu\Omega$ at $8\,\text{T}$ (Fig.~\ref{fig:2}b, inset). The fitted zero-field intercept ($84.12 \pm 0.12\,\mu\Omega$) represents a $Q_L$-referenced effective baseline containing field-independent antenna, dielectric, and holder losses, rather than the intrinsic residual film resistance. Conversely, the perpendicular orientation ($B \perp ab$-plane) represents the most dissipative geometry. To immobilize perpendicular fluxoids, the EuBCO tapes incorporate $c$-axis aligned artificial pinning centers (APCs)~\cite{Romanov2020HighFreq}.

To model the high-field vortex dynamics, we evaluated the field-swept loaded quality factor $Q_L(B)$ of the rutile resonator within the framework of the Gittleman-Rosenblum model~\cite{Gittleman1966VortexModel, Golosovsky1996YBCOvorexRev}:
\begin{equation}
R_{s,\text{fit}}(B) = R_{s0} + \sqrt{\frac{\mu_0 \omega}{2}} \left[ \sqrt{(a B)^2 + (b + c B)^2} - (b + c B) \right]^{1/2}
\end{equation}
where $\mu_0 = 4\pi \times 10^{-7}\,\text{H/m}$, $\omega = 2\pi\nu$ is the operational angular frequency with the measured cyclic frequency $\nu$ expressed in Hz, and $a, b, c$ are phenomenological coefficients parameterizing the magnetic-field dependence of vortex dissipation. High-field fits using ramping data above $1\,\text{T}$ (perpendicular) and $2\,\text{T}$ (parallel) constrain the perpendicular field-induced increase to $\Delta R_s^\perp = 179.84 \pm 0.90\,\mu\Omega$ at $8\,\text{T}$ ($R^2 = 0.9848$; Fig.~\ref{fig:2}b, inset). These fit uncertainties represent formal fit-covariance without including geometry-factor or system-systematic uncertainties.

In light of established vortex-matter phase studies in REBCO coated conductors~\cite{Blatter1994VortexRev}, the non-monotonic behavior observed below $0.5\,\text{T}$ is consistent with initial flux penetration or vortex-state evolution, although instrumental or mode-structure effects are alternative contributing factors. Additionally, parallel dataset points between $1.09\,\text{T}$ and $1.58\,\text{T}$ adjacent to a disconnected mode-crossing region were omitted from the fitted dataset.

Figure~\ref{fig:2}b (main panel) displays the frequency dependence of $R_s$ and $\Delta R_s$ at $8\,\text{T}$. While polycrystalline copper follows a power law ($R_s^{\text{Cu}} \propto \nu^{0.59}$), the dashed $\nu^2$ curves represent theoretical guide lines anchored to the rutile measurements ($\Delta R_s^\parallel = 5.13\,\mu\Omega$ and $\Delta R_s^\perp = 179.84\,\mu\Omega$ near $8\,\text{GHz}$). For the $5.40\,\text{GHz}$ 3D cavity, joint-fit analysis yields effective assembled-cavity losses of $31.36\,\mu\Omega$ at $8\,\text{T}$ and $22.3\,\mu\Omega$ at $0\,\text{T}$. Their difference, representing the field-induced increase $\Delta R_s(8\,\text{T}) = 9.06\,\mu\Omega$ (purple diamond; Fig.~\ref{fig:2}b), naturally isolates the magnetic dissipation by canceling field-independent background losses, placing the 3D cavity performance squarely within these anisotropic material bounds.

\begin{figure}[htbp]
    \centering
    \includegraphics[width=0.85\textwidth]{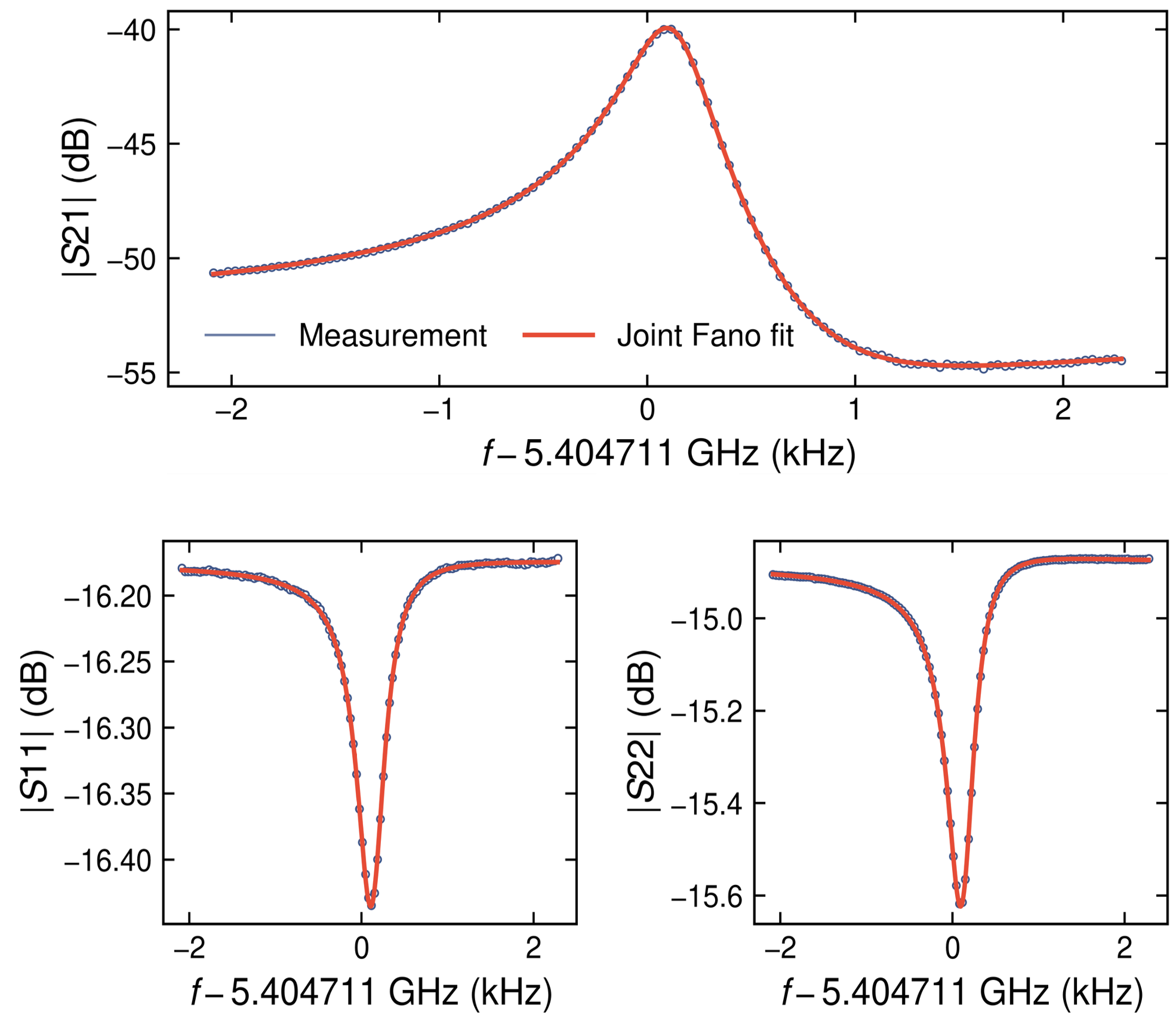}
    \caption{\textbf{Joint scalar resonance fit for the $5.40\,\text{GHz}$ EuBCO cavity at $8\,\text{T}$.} Upper panel: $|S_{21}|$ transmission spectrum. Lower panels: $|S_{11}|$ and $|S_{22}|$ reflection spectra. Blue open circles represent the measured logarithmic magnitude data ($f_c \approx 5.404711\,\text{GHz}$, span = $4.37\,\text{kHz}$, IFBW = $1\,\text{kHz}$). Red curves denote the simultaneous non-linear least-squares Fano fit described in the Supplementary Note. The joint extraction yields $f_0 \approx 5.404711\,\text{GHz}$, pole linewidth $\Gamma \approx 401\,\text{Hz}$, $Q_L \approx 1.35 \times 10^7$, and positive-reflection couplings $\beta_1 \approx 0.016$ and $\beta_2 \approx 0.042$, resulting in a derived unloaded quality factor $Q_0 \approx 1.42 \times 10^7$.}
    \label{fig:SI_fano}
\end{figure}

Crucially, the ultimate macroscopic validation of this conductive-backed architecture lies in the resonant performance of the fully assembled 3D cavities. As shown in Fig.~\ref{fig:2}a, all conductive-backed prototypes exhibit exceptional high-field resilience compared to standard copper ($Q \lesssim 10^5$). Leveraging the APC-doped EuBCO film within Architecture A, the $2.27\,\text{GHz}$ prototype achieved $Q_0 \approx 3.5 \times 10^6$ at $8\,\text{T}$, alongside a $2.30\,\text{GHz}$ GdBCO variant maintaining $Q_0 \approx 5 \times 10^5$, while the stiffness-optimized $5.40\,\text{GHz}$ cavity reached a saturated quality factor of $Q_0 = 1.4 \times 10^7$ at $8\,\text{T}$ (Fig.~\ref{fig:2}a, Table~\ref{tab:cavities}, and Supplementary Fig.~\ref{fig:SI_fano}). 

Building upon this high-field performance, specialized Architecture A cavities were successfully deployed in dark matter searches: the $2.27\,\text{GHz}$ EuBCO cavity ($Q_0 \approx 3.5 \times 10^6$) drove a search for axion quark nuggets~\cite{AQN_PRD_2025}, while the $5.40\,\text{GHz}$ cavity ($Q_0 = 1.4 \times 10^7$) enabled a search operating in the ultra-high-$Q$ regime where the cavity linewidth is narrower than the predicted axion signal linewidth ($\Delta\nu_c < \Delta\nu_a$) (manuscript in preparation).

\subsection*{Axion dark matter search with an operational HTS haloscope}

In the CAPP-PACE haloscope, maximizing the axion scanning rate ($d\nu/dt \propto V^2 Q_L T_{\text{sys}}^{-2}$) requires balancing the active volume ($V$) against the quality factor ($Q_L$). Therefore, we deployed the volume-optimized $2.3\,\text{GHz}$ GdBCO pathfinder cavity (Architecture B; No.~2 in Table~\ref{tab:cavities}). Unlike the aluminum-wedge design, this monolithic inner-mold copper-frame architecture maximizes the active volume ($V = 1.5\,\text{L}$) within the magnet bore while guaranteeing robust millikelvin thermalization.

\begin{figure}[htbp]
    \centering
    \includegraphics[width=\textwidth]{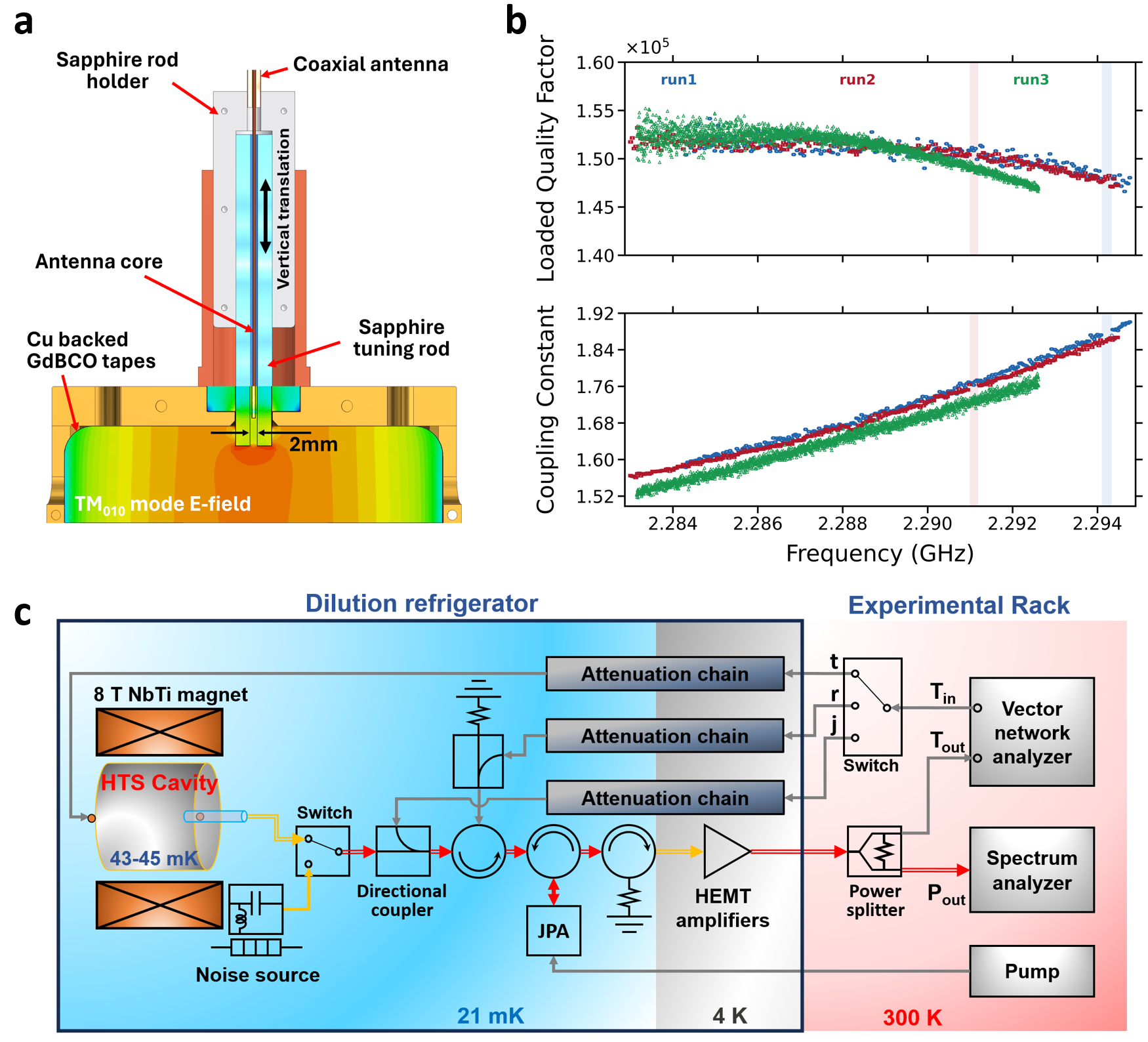}
    \caption{\textbf{CAPP-PACE detector integration and cryogenic performance.} 
    \textbf{(a)} Cross-sectional schematic of the non-contact axial tuning mechanism overlaid with the $TM_{010}$ electric field distribution simulated via 3D electromagnetic eigenmode solver (CST Microwave Studio (MWS), within CST Studio Suite 2015), showing field concentration inside the $9.7\,\text{mm}$ diameter sapphire rod ($\epsilon_r \approx 10$) with its $2\,\text{mm}$ central hole.
    \textbf{(b)} Real-time \textit{in-situ} measurements of the operational loaded quality factor ($Q_L$) and antenna coupling strength ($\beta$) recorded at each tuning step during live physics data acquisition as a function of resonant frequency at $B_{\text{RMS}} = 6.95\,\text{T}$ ($B_{\text{max}} = 8.0\,\text{T}$) and cavity temperature $T_{\text{cav}} = 45\,\text{mK}$. Loaded quality factor $Q_L$ remains $\sim 5 \times$ higher than the conventional copper baseline ($Q_L^{\text{Cu}} \approx 3.0 \times 10^4$) across the entire tuning band. Shaded vertical bands mark unmeasured frequency intervals resulting from localized tuning mechanics in Run~1 (blue) and Run~2 (red).
    \textbf{(c)} Schematic of the microwave receiver chain and experimental readout layout (adapted from Refs.~\cite{Kwon2021FirstResults, Kim2023NearQuantum}). The Architecture B HTS cavity is thermalized at $43\text{--}45\,\text{mK}$ inside an $8.0\,\text{T}$ NbTi magnet bore, connected via a cold RF switch to a calibrated noise source for Y-factor noise calibration. The signal path routes through a directional coupler and flux-driven JPA at the $21\,\text{mK}$ mixing chamber plate, two-stage high electron mobility transistor (HEMT) amplifiers at $4\,\text{K}$, and room-temperature ($300\,\text{K}$) vector network analyzer (VNA) and spectrum analyzer (SA) instruments for live data acquisition.}
    \label{fig:3}
\end{figure}

Dynamic frequency tuning is essential for continuous haloscope scanning. To tune the pathfinder without mechanical friction on the indium-welded seams, we developed an axial non-contact tuning mechanism featuring a $9.7\,\text{mm}$ diameter cryogenic sapphire rod inserted through the top aperture. This rod incorporates a coaxial $2\,\text{mm}$ central hole that accommodates the stationary antenna core without contact (Fig.~\ref{fig:3}a). As the high-dielectric sapphire rod ($\epsilon_r \approx 10$) inserts vertically along the axis, the $TM_{010}$ electric field progressively concentrates inside the dielectric region (Fig.~\ref{fig:3}a), setting the operational form factor to $C_{010} \approx 0.58\text{--}0.62$ across the tuning band (simulated via COMSOL Multiphysics; Table~\ref{tab:runs}).

Simultaneously, \textit{in-situ} characterization during live physics data acquisition confirmed that the loaded quality factor ($Q_L \approx 1.47\text{--}1.55 \times 10^5$; Fig.~\ref{fig:3}b) and antenna coupling coefficient ($\beta \approx 1.5\text{--}1.9$; Fig.~\ref{fig:3}b), recorded in real time at every tuning step, remained highly stable throughout all three physics runs. Evaluated step-by-step via $Q_0 = Q_L(1+\beta)$, these parameters yield an overall run-averaged RMS unloaded quality factor of $Q_{0,\text{RMS}} \approx 4.05 \times 10^5$ (reaching $4.10 \times 10^5$ for Run~1). This expected reduction from the empty-cavity baseline ($Q_0 \approx 5.0 \times 10^5$, measured without the tuning rod) is primarily attributed to electric field concentration interacting with the normal-conducting boundaries of the rod insertion port. Relative to the $Q_L^{\text{Cu}} = 3.00 \times 10^4$ baseline used for the 2020 copper comparator in Table~\ref{tab:comparison}, the operational loaded-$Q$ enhancement is approximately five-fold. The low loss tangent of cryogenic sapphire makes its intrinsic dielectric contribution subdominant in the overall loss budget.

Thermalizing HTS materials at millikelvin temperatures poses a major engineering challenge due to their inherently low out-of-plane thermal conductivity. We resolved this by utilizing the continuous copper stabilizer and the external copper frame as primary thermal conduction paths. Additionally, all copper and brass support structures were chemically etched to strip away thermally resistive oxide layers prior to assembly. This thermal engineering drove the cavity's physical temperature down to a stable $43\text{--}45\,\text{mK}$, while maintaining the flux-driven Josephson Parametric Amplifier (JPA) at $21\,\text{mK}$ on the mixing chamber plate within the complete cryogenic receiver chain (Fig.~\ref{fig:3}c)~\cite{Kim2023NearQuantum}.

Under these thermalized operating conditions, we quantified the receiver chain's performance via a two-step noise verification procedure combining the Y-factor method and the Spectrum Comparison Method (SCM)~\cite{Kwon2021FirstResults, Kim2023NearQuantum}. Through this dual-verification (detailed in Methods), we confirmed a stable system noise temperature of $T_{\text{sys}} \approx 175\text{--}185\,\text{mK}$ across the entire tuning range (Table~\ref{tab:runs}). Approaching the quantum noise threshold ($h\nu/k_B \approx 110\,\text{mK}$ at $2.3\,\text{GHz}$), this low noise figure minimizes power spectral fluctuations and directly accelerates the experimental scanning speed.

\begin{figure}[htbp]
    \centering
    \includegraphics[width=\textwidth]{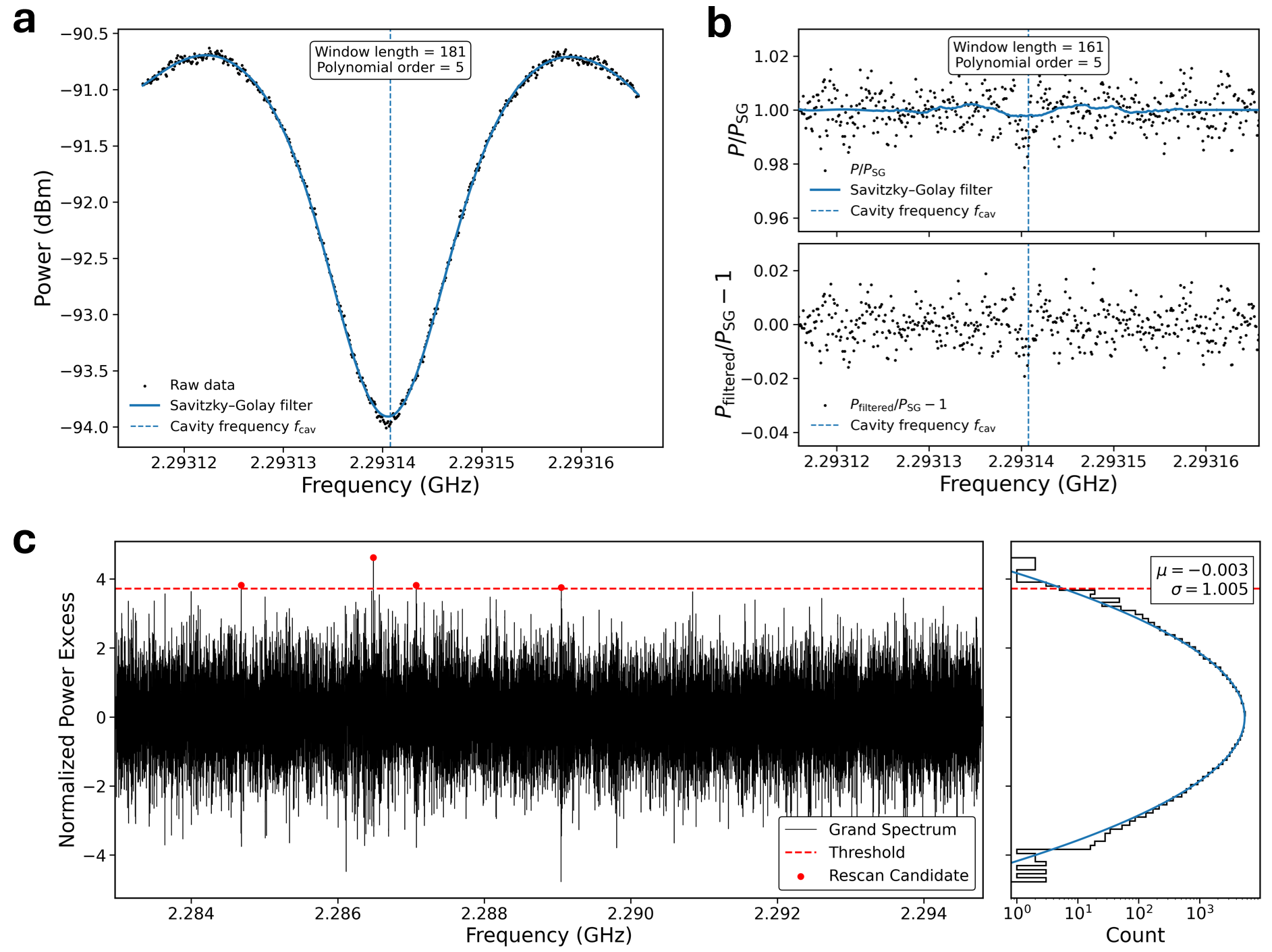}
    \caption{\textbf{Two-stage baseline removal and grand spectrum candidate evaluation.} 
    \textbf{(a)} Representative raw intermediate-frequency power spectrum overlaid with the first-stage Savitzky-Golay (SG) baseline fit (blue solid line; window length = 181, polynomial order = 5). The vertical dashed line marks the cavity resonant frequency ($f_{\text{cav}}$). 
    \textbf{(b) Upper:} Intermediate spectrum after first-stage baseline division ($P/P_{\text{SG}}$) alongside the second-stage SG filter (blue solid line; window length = 161). \textbf{Lower:} Final normalized power excess residuals ($P_{\text{filtered}}/P_{\text{SG}} - 1$) displaying a flattened, zero-mean noise baseline. 
    \textbf{(c) Left:} Grand combined power spectrum plotted against frequency/axion mass across the scanning range, displaying the predefined candidate selection threshold ($3.718\sigma$; red dashed line) and identified rescan candidates (red dots). \textbf{Right:} Histogram of normalized power excess from the grand spectrum fitted with a Gaussian distribution (blue curve), confirming ideal Gaussian noise characteristics ($\mu = -0.003, \sigma = 1.005$).}
    \label{fig:4}
\end{figure}

Leveraging this low-noise HTS haloscope, we executed a dark matter search, collecting power spectra across the $2.282\text{--}2.295\,\text{GHz}$ frequency range (axion mass $m_a \in [9.4376, 9.4914]\,\mu\text{eV}$; see Table~\ref{tab:runs} for run parameters). To isolate potential axion conversion signals, we processed the raw spectra by systematically removing frequency-dependent baseline variations using a two-stage Savitzky-Golay filter (Fig.~\ref{fig:4}a,b). As optimized via Monte Carlo simulations (detailed in Methods), this strategy successfully maintained an effective signal-retention efficiency of $\eta_{\text{BL}} \approx 52\%$ ($\eta_{\mathrm{BL}}^{\mathrm{HTS}} = 0.517$).

During the candidate search, five spectral peaks exceeded the candidate threshold ($3.718\sigma$). Each candidate was subjected to a 10-hour re-scan procedure; none persisted with the expected signal behavior and all were therefore rejected as candidates. This outcome is consistent with statistical or transient noise but does not identify a unique physical noise mechanism. As shown in the grand combined spectrum (Fig.~\ref{fig:4}c, left), the background noise distribution is well described by a Gaussian profile ($\mu = -0.003, \sigma = 1.005$; Fig.~\ref{fig:4}c, right).

\begin{figure}[htbp]
    \centering
    \includegraphics[width=0.8\textwidth]{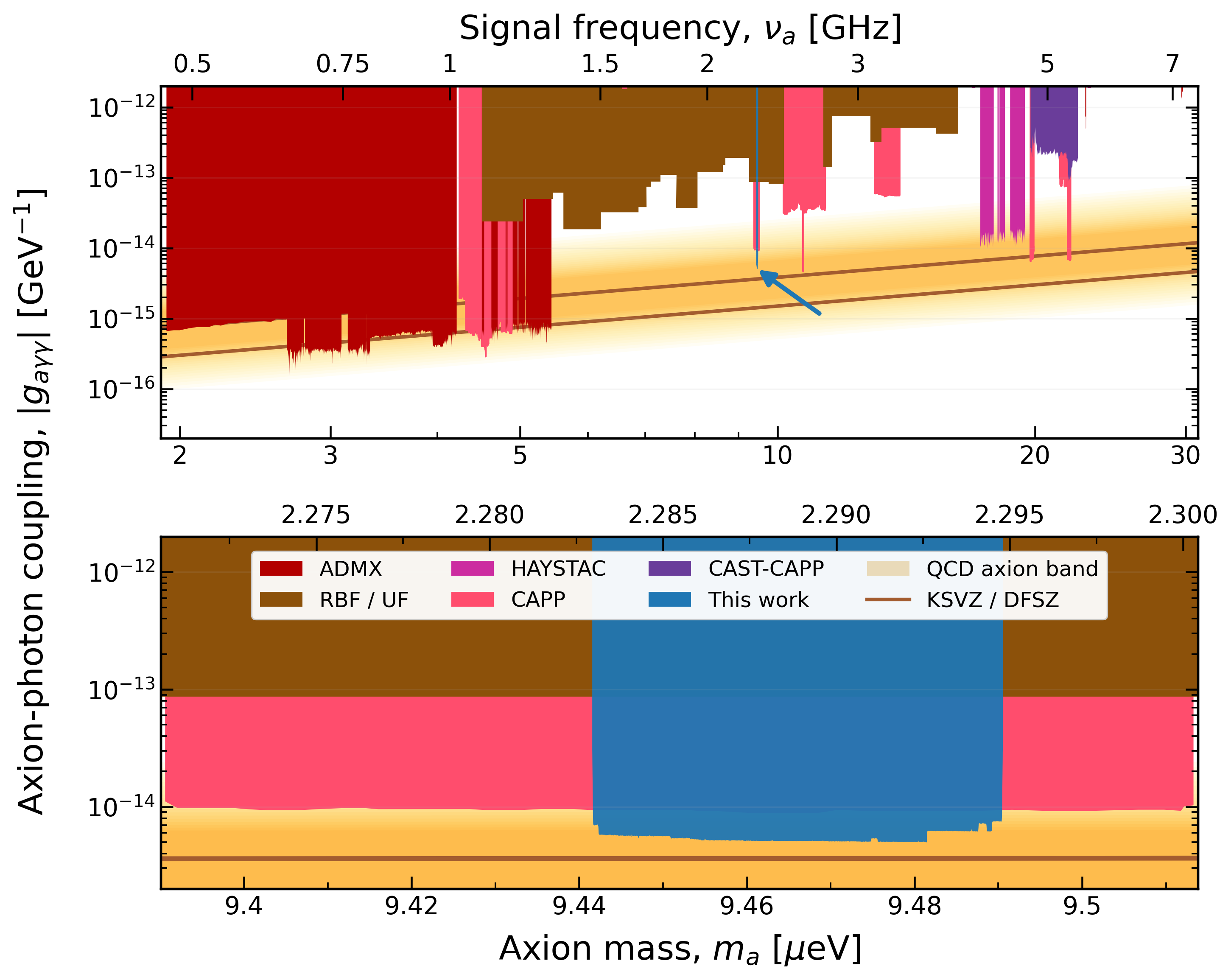}
    \caption{\textbf{Axion-photon coupling exclusion limits.} 
    \textbf{Top panel:} Global axion dark matter parameter space ($|g_{a\gamma\gamma}|$ versus axion mass $m_a$ and signal frequency $\nu_a$) contextualizing this work (blue region, indicated by the arrow) against major existing haloscope search results (ADMX, RBF/UF, HAYSTAC, CAPP, CAST-CAPP) and theoretical QCD axion models (KSVZ/DFSZ benchmark lines and yellow band). The background exclusion limits and theoretical model bands are adapted from the open-source AxionLimits database~\cite{AxionLimits}.
    \textbf{Bottom panel:} Detailed $90\%$ confidence level (C.L.) upper limit on $|g_{a\gamma\gamma}|$ established in this work across the mass range $m_a \in [9.4376, 9.4914]\,\mu\text{eV}$ ($\nu_a \in [2.282, 2.295]\,\text{GHz}$). The CAPP-PACE HTS haloscope deployment achieved a physical sensitivity reaching $1.4 \times |g_{a\gamma\gamma}^{\text{KSVZ}}|$, delivering an $8.3$-fold scan-rate advantage over conventional copper haloscopes. The characteristic staircase structure along the lower exclusion boundary reflects non-uniform cumulative integration time across overlapping physics runs and localized piezo-actuator tuning steps.}
    \label{fig:5}
\end{figure}

Consequently, we established a $90\%$~confidence-level (C.L.) upper limit on the axion-photon coupling constant ($\vert{}g_{a\gamma\gamma}\vert{}$) across the scanned mass range $m_a \in [9.4376, 9.4914]\,\mu\text{eV}$ (Fig.~\ref{fig:5}). Figure~\ref{fig:5} (top panel) contextualizes our result within the global dark matter parameter space, while Figure~\ref{fig:5} (bottom panel) provides a detailed view of the excluded coupling region. The resulting exclusion limit displays a characteristic stepped (staircase) lower boundary. This step structure arises from the non-uniform cumulative integration time across the band, as the three independent physics runs (Table~\ref{tab:runs}) covered overlapping yet slightly distinct frequency sub-intervals. Furthermore, stick-slip friction during the vertical actuation of the heavy sapphire rod assembly by the piezo motor in Runs~1 and 2 occasionally caused sudden tuning steps, leaving narrow unmeasured frequency gaps in individual runs (Fig.~\ref{fig:3}b). Spectral bins combining integration time from all three runs achieved deeper coupling sensitivity compared to bins covered by one or two runs, reaching a physical sensitivity of $1.4 \times |g_{a\gamma\gamma}^{\text{KSVZ}}|$. The resulting exclusion limits are the most stringent to date in this mass range at 90\% confidence level.

To quantify how this HTS haloscope deployment translates into scanning efficiency gains, Table~\ref{tab:comparison} presents a comprehensive 1:1 parameter and scan-rate comparison against our baseline 2020 copper haloscope~\cite{Kim2023NearQuantum}. By implementing the volume-efficient delaminated HTS structure, we achieved an operational loaded quality factor 5 times higher than copper ($Q_L \approx 1.51 \times 10^5$ versus $Q_L^{\text{Cu}} \approx 3.00 \times 10^4$) while securing a larger active volume ($V = 1.50\,\text{L}$ versus $1.12\,\text{L}$) and a lower effective system noise temperature ($T_{\text{sys}} \approx 177\,\text{mK}$ versus $234\,\text{mK}$). Evaluating the full radiometer equation ($d\nu/dt \propto V^2 B^4 C_{010}^2 Q_L T_{\text{sys}}^{-2}$), this combination yields an ideal hardware scan-rate enhancement of $20.0$-fold over copper. Factoring in baseline retention efficiencies [$\eta_{\mathrm{BL}}^{\mathrm{HTS}}=0.517$ vs. $\eta_{\mathrm{BL}}^{\mathrm{Cu}}=0.80$, scaling as $(\eta_{\mathrm{BL}})^2$], this translates to an analysis-adjusted enhancement of $20.0 \times (0.517/0.80)^2 \approx 8.3$-fold. This combined hardware and operational speedup confirms that strain-controlled conductive-backed HTS cavities dramatically accelerate the search for QCD axion dark matter.

\section*{Discussion}

Demonstrating an unloaded quality factor exceeding $1.4 \times 10^7$ at $8\,\text{T}$ in a fixed-frequency EuBCO cavity, alongside an operational near-quantum-noise axion search yielding an $8.3$-fold scan-rate enhancement via a tunable GdBCO haloscope, establishes substrate-stripped conductive-backed HTS cavities as a versatile platform for high-field physics. These complementary demonstrations highlight the adaptability of the architecture across distinct operational constraints. More recently, our group successfully applied this substrate-stripped REBCO framework to a broadly tunable copper-shell cavity for axion searches in the $15.30\text{--}15.85\,\mu\text{eV}$ mass range~\cite{Lee2026TunableHTS}.

As cavity quality factors enter the ultra-high-$Q$ regime ($Q_0 > 10^7$, Architecture A), operating near or exceeding the axion linewidth limit ($Q_L \gtrsim Q_a \approx 10^6$) introduces novel signal-processing considerations. When the cavity linewidth becomes narrower than the predicted axion signal ($\Delta\nu_c \le \Delta\nu_a$), conventional baseline filters risk shaving off genuine axion conversion power. While our dual-pass Savitzky-Golay filter preserved an effective signal retention efficiency of $51.7\%$ in the operational GdBCO haloscope ($Q_L \approx 1.5 \times 10^5$), deploying ultra-high-$Q$ cavities encourages alternative baseline strategies. Future high-$Q$ searches can leverage wideband Josephson Parametric Amplifiers (JPAs)~\cite{BroadbandJPA2019APL} or Traveling Wave Parametric Amplifiers (TWPAs)~\cite{Planat2020PRX, DiVora2023PRD}. Because the cavity is tuned in fine steps much smaller than the amplifier bandwidth, the background transfer function remains virtually static across adjacent tuning steps. Implementing multi-step joint baseline fitting or inter-step template subtraction could decouple the static amplifier profile without fitting out the stationary axion signal, substantially mitigating signal attenuation penalties.

Addressing these signal-processing considerations alongside our structural progression from a $2.27\,\text{GHz}$ prototype ($Q_0 \approx 3.5 \times 10^6$) to a stiffness-optimized $5.40\,\text{GHz}$ cavity ($Q_0 = 1.4 \times 10^7$) directly informs the scalability roadmap for upcoming flagship searches, such as the Dark Matter Axion Group (DMAG)/CAPP-12TB experiment~\cite{yi2023axion,ahn2024prx} utilizing a $12\,\text{T}$ magnetic field and a $32\,\text{cm}$ large-bore magnet. Ideal electromagnetic boundary simulations demonstrated that $TM_{010}$ quality factors of order $10^7\text{--}10^8$ are achievable for longitudinal gaps up to $100\,\mu\text{m}$ under ideal conductor conditions~\cite{ahn2022biaxially}. By leveraging modern precision machining complying with standard IT6--IT7 industrial grades~\cite{ISO286} alongside thermal-expansion-matched frames, constructing large-volume ($> 30\,\text{L}$) high-$Q$ HTS cavities appears technically promising. While practical cryogenic assembly will require further engineering validation regarding cumulative seam alignment and thermal deformation, conductive-backed HTS cavities establish a viable path toward scaling high-$Q$ resonators for large-scale dark matter searches.

Beyond dark matter searches, high-$Q$ HTS cavities operating in strong magnetic fields open opportunities in accelerator RF technology and hybrid quantum systems. REBCO coatings have been investigated for FCC-hh beam screens under $16\text{-T}$-class fields~\cite{Romanov2020HighFreq}, while muon-collider cooling channels require RF structures to operate inside multi-tesla solenoids~\cite{Bowring2020PRAB}. The substrate-stripped, copper-backed resonators demonstrated here provide an architecture for exploring low-loss microwave structures in such magnetic environments. Furthermore, high-$Q$ HTS cavities could provide a three-dimensional microwave environment for future hybrid circuit-QED experiments with magnetic-field-compatible transmon qubits~\cite{Krause2022MagneticTransmon,Xiang2013Hybrid}. Ultimately, strain-controlled conductive-backed HTS cavities stand as a scalable platform, primed to advance sensitivity frontiers in both fundamental particle physics and quantum technology.

\section*{Methods}

\subsection*{Cavity Design and Fabrication}
The high-temperature superconducting (HTS) cavities were constructed using commercially available biaxially-textured REBCO coated conductors (Table~\ref{tab:tapes}). The GdBCO tapes (Tape~1; THEVA
Dünnschichttechnik GmbH, Ismaning, Germany) featured a $100\,\mu\text{m}$ copper stabilizer, a $1.5\,\mu\text{m}$ Ag protection layer, a $3\text{--}5\,\mu\text{m}$ HTS layer, a $3.5\,\mu\text{m}$ MgO buffer layer, and a $50\,\mu\text{m}$ Hastelloy substrate. The EuBCO tapes (Tape~2; Fujikura Ltd., Tokyo,
Japan), doped with BaHfO$_3$ artificial pinning centers, comprised a $20\,\mu\text{m}$ copper stabilizer, a $2\,\mu\text{m}$ Ag layer, a $2.5\,\mu\text{m}$ HTS layer, a $0.7\,\mu\text{m}$ MgO buffer, and a $50\,\mu\text{m}$ Hastelloy substrate.

To suppress cross-seam RF leakage while preserving film performance, we implemented a controlled mechanical delamination process. The tape was guided around a cylindrical jig with a bending radius of $R = 20\,\text{mm}$ and peeled at an angle of $\theta \approx 90^\circ$. While this geometry concentrates opening-type stress at the peel front, actual peel forces are governed by a complex combination of interfacial fracture energy and plastic dissipation within the copper stabilizer~\cite{Hojo2012ModeI, Lu2025Peel}. Furthermore, because the neutral-axis position in a multi-material composite tape shifts according to layer-dependent elastic moduli, estimating HTS layer strain purely from total-thickness bending approximations ($t/2R$) can be imprecise. We therefore validated the process through direct functional microwave benchmarking: achieving a high-field quality factor ($Q_0 = 1.4 \times 10^7$ at $8\,\text{T}$) demonstrates preservation of high-frequency superconducting performance after full removal of the $50\,\mu\text{m}$ Hastelloy substrate.

Depending on the architecture (Table~\ref{tab:cavities}), the delaminated segments were attached to precision Aluminum 6061 wedges with joint faces co-polished for sub-10-$\mu\text{m}$ alignment (Architecture A), or aligned over an inner cylindrical mold to form two semi-cylindrical shell halves, secured from the outside via low-temperature indium cold-welding, and mounted into an outer structural frame (Architecture B). Mechanical integrity and outer conductive shielding behind the assembly seams were established utilizing the robust copper stabilizer backing. To ensure optimal thermalization within the dilution refrigerator, all copper and brass support structures were chemically etched using Citranox acidic cleaner (Alconox) before assembly, minimizing boundary thermal resistance.

For the assembled haloscope cavities, the axion-photon conversion form factor, $C_{010}$, incorporating the spatial dielectric energy distribution of the sapphire tuning rod ($\epsilon_r \approx 10$), is defined as:
\begin{equation}
C_{010} = \frac{\left| \int_V \mathbf{E}_{010}(\mathbf{r}) \cdot \mathbf{B}_0(\mathbf{r}) \, dV \right|^2}{B_{\text{RMS}}^2 V \int_V \epsilon_r(\mathbf{r}) |\mathbf{E}_{010}(\mathbf{r})|^2 \, dV}
\end{equation}
where $\mathbf{B}_0(\mathbf{r})$ is the static magnetic-field map and $B_{\text{RMS}} \equiv \sqrt{V^{-1}\int_V |\mathbf{B}_0(\mathbf{r})|^2\,dV}=6.95\,\text{T}$ is its volume-averaged root-mean-square magnitude over the active volume $V=1.5\,\text{L}$ at a peak central field $B_{\text{max}}=8.0\,\text{T}$. To achieve near-optimal scanning speed ($d\nu/dt \propto \beta^2/(1+\beta)^3$), the cavity-to-receiver coupling was dynamically maintained in the overcoupled regime ($\beta \approx 1.5\text{--}1.9$), yielding $>97\%$ of the theoretical maximum rate.

\subsection*{Quality Factor and Surface Resistance Measurements}
The high-field surface resistance ($R_s$) of the REBCO films was characterized using a modified Hakki-Coleman dielectric resonator setup operating in the $TE_{011}$ mode. The fixture comprises a rutile ($\mathrm{TiO}_2$) crystal cylinder (height $3\,\text{mm}$, radius $2\,\text{mm}$, temperature-saturated sub-kelvin dielectric constant $\epsilon_r \approx \mathcal{O}(100)$) enclosed within a copper cavity with an inner diameter of $6\,\text{mm}$, defining a resonant frequency near $8\,\text{GHz}$. The $TE_{011}$ mode generates purely azimuthal surface currents ($\mathbf{J}_\phi$) on the bounding sample surfaces, ensuring zero current flow across mechanical seams ($\mathbf{J}_{\text{joint}} = 0$) and effectively eliminating contact loss at the conductor-holder boundaries. Generally, the relation between a resonator's quality factor and the surface resistance is governed by the electromagnetic geometry factor, defined as:
\begin{equation}
G \equiv \frac{\omega \mu_0 \int_V |\mathbf{H}|^2 dV}{\int_S |\mathbf{H}_t|^2 dS}
\end{equation}
For this rutile fixture containing two nominally identical REBCO tape endplates, 3D electromagnetic simulations using COMSOL Multiphysics version 5.2a (COMSOL AB) yield a combined geometry factor of $G_{\text{rutile}} = 104.8\,\Omega$ for the sample pair.

The total surface resistance, $R_s$, of a Type-II superconductor in the mixed state is modeled as the sum of Cooper pair, vortex motion, and residual resistance contributions:
\begin{equation}
R_s(\nu, T, B) = R_{\text{BCS}}(\nu, T, 0) + R_v(\nu, T, B) + R_{\text{res}}(\nu, T, B)
\end{equation}
In the low-frequency pinned regime below the depinning frequency, vortex motion follows the Gittleman-Rosenblum model, describing a forced damped oscillator incorporating the Lorentz force ($\mathbf{F}_L = \mathbf{J} \times \mathbf{\Phi}_0$), the pinning force ($\mathbf{F}_p = -k_p \mathbf{u}$), and the drag force ($\mathbf{F}_v = -\eta \dot{\mathbf{u}}$):
\begin{equation}
\eta \dot{\mathbf{u}} + k_p \mathbf{u} = \mathbf{J} \times \mathbf{\Phi}_0 \hat{n}_v
\end{equation}
where $\mathbf{u}$ is the vortex displacement vector, $\mathbf{\Phi}_0$ is the magnetic flux quantum ($\Phi_0 = h/2e \approx 2.07 \times 10^{-15}\,\text{Wb}$), $k_p$ is the Labusch pinning parameter, $\eta$ is the vortex viscosity, $\omega = 2\pi\nu$ represents the operational angular frequency corresponding to the measured cyclic frequency $\nu$, and $\hat{n}_v$ is the unit vector along the vortex line. Extrapolating reported depinning-frequency fits for commercial coated conductors~\cite{Romanov2020HighFreq} toward the low-temperature limit yields the model input estimate $\omega_p/2\pi \simeq 120\,\text{GHz}$. The complex vortex resistivity ($\rho_v$) and surface impedance ($Z_v = R_v + i X_v$) are expressed as:
\begin{equation}
\rho_v = \frac{B\Phi_0}{\eta} \cdot \frac{i\omega/\omega_p}{1 + i\omega/\omega_p}
\end{equation}
\begin{equation}
Z_v = R_v + iX_v = \sqrt{i\omega\mu_0\rho_v}
\end{equation}

Assuming antenna, dielectric, and sidewall losses remain field-independent, the field-induced surface resistance variation of the tape samples is extracted via:
\begin{equation}
\Delta R_s(B) = G_{\text{rutile}} \left[ \frac{1}{Q_0(B)} - \frac{1}{Q_0(0)} \right] = G_{\text{rutile}} \left[ \frac{1}{Q_L(B)} - \frac{1}{Q_L(0)} \right]
\end{equation}
Because the simulated $G_{\text{rutile}}$ already accounts for both sample surfaces, no additional factor of two is applied. Absolute values proportional to $G_{\text{rutile}}/Q_0$ retain an unknown field-independent offset and serve as effective fit intercepts. Field-induced variations were fitted using the nonlinear Gittleman-Rosenblum physical model to evaluate vortex dissipation.

For macroscopic 3D cavities operating in the $TM_{010}$ mode, the measured total loss is equivalently expressed as an effective surface resistance:
\begin{equation}
R_{s,\text{eff}}(B) \equiv \frac{G_{\text{cavity}}}{Q_0(B)}
\end{equation}
For the $5.40\,\text{GHz}$ EuBCO cavity (No.~4), analogous COMSOL simulations yield $G_{\text{cavity}} = 446.7\,\Omega$. The joint scalar Fano fit at $8\,\text{T}$ gives $Q_0 \approx 1.423\times 10^7 $, resulting in $R_{s,\text{eff}}(8\,\text{T}) = 31.36\,\mu\Omega$. Subtracting the measured zero-field baseline $R_{s,\text{eff}}(0\,\text{T}) = 22.30\,\mu\Omega$ strictly cancels field-independent seam and test-cell losses, isolating the field-induced surface resistance increase:
\begin{equation}
\Delta R_s(8\,\text{T}) = G_{\text{cavity}} \left[ \frac{1}{Q_0(8\,\text{T})} - \frac{1}{Q_0(0\,\text{T})} \right] = 31.36 - 22.30 = 9.06\,\mu\Omega
\end{equation}
Thus, the reported $9.06\,\mu\Omega$ rigorously represents the isolated field-induced surface resistance increase between $0$ and $8\,\text{T}$, rather than an absolute uncalibrated intrinsic film resistance.

Resonance quality factors were extracted in the frequency domain using a vector network analyzer (VNA; N5232A, Keysight Technologies, Santa Rosa, CA, USA) via joint scalar Fano resonance fitting across $|S_{21}|$, $|S_{11}|$, and $|S_{22}|$ (Supplementary Fig.~\ref{fig:SI_fano}). For the $5.40\,\text{GHz}$ cavity at $8\,\text{T}$, the measurement span of $4.37\,\text{kHz}$ was sampled with 1,001 points ($\sim 4.37\,\text{Hz}$ bin spacing), yielding a fitted loaded pole linewidth of $f_0/Q_L = 401.38 \pm 0.13\,\text{Hz}$ and an unloaded linewidth of $f_0/Q_0 = 379.46\,\text{Hz}$. Narrow intermediate frequency bandwidths (IFBW) and extended sweep dwell times ensured steady-state energy equilibrium during resonance sweeps.

\subsection*{Experimental Setup and Data Acquisition}
The pathfinder HTS cavity (Architecture B) was thermalized to the mixing chamber plate of a dilution refrigerator (LD400, Bluefors, Helsinki, Finland) and operated inside a $12\,\text{cm}$ bore of $8\,\text{T}$ superconducting magnet (American Magnetics, Oak Ridge, TN, USA) with a peak central field $B_{\text{max}} = 8.0\,\text{T}$ (volume-averaged $B_{\text{RMS}} = 6.95\,\text{T}$ over $V = 1.5\,\text{L}$; Fig.~\ref{fig:3}c). The cryogenic receiver chain, microwave amplification stages, and thermal anchoring layout were configured as shown in Fig.~\ref{fig:3}c, following our established setup detailed in Ref.~\cite{Kim2023NearQuantum}. Dynamic frequency tuning was executed using a cryogenic linear piezo actuator (ANPz101eXT12/RES/LT/HV, attocube systems, Haar, Germany) that drove the sapphire rod assembly vertically through the top insertion aperture, with the corresponding electric-field distribution shown in Fig.~\ref{fig:3}a calculated using the eigenmode solver in CST Microwave Studio (MWS), CST Studio Suite 2015.

Data acquisition was conducted across three independent physics runs (Runs~1, 2, and 3; detailed in Table~\ref{tab:runs}), achieving a net cumulative live integration time of approximately 30 days after accounting for cryogenic cycling, scheduled maintenance, and tuning overheads. At each frequency tuning step ($\Delta\nu \approx 4\,\text{kHz}$), a tightly interleaved measurement routine was executed. First, a VNA performed \textit{in-situ} measurements of complex reflection ($S_{11}$) and transmission ($S_{21}$) spectra to track the cavity resonant frequency ($f_0$), operational loaded quality factor ($Q_L \approx 1.47\text{--}1.55 \times 10^5$), and antenna coupling coefficient ($\beta \approx 1.52\text{--}1.88$; Fig.~\ref{fig:3}b). Subsequently, the microwave pump generator for the flux-driven Josephson Parametric Amplifier (JPA) was automatically adjusted to align the JPA gain profile ($G_J \approx 21\,\text{dB}$) with the cavity resonance. Finally, power spectra were digitized and averaged in 1-second intervals using a spectrum analyzer (FSV7, Rohde \& Schwarz, Munich, Germany), yielding 180 to 520 individual spectra per tuning step across the three physics runs (Table~\ref{tab:runs}).

\subsection*{Noise Measurement}
To characterize the system noise temperature ($T_{\text{sys}}$), we employed the two-step verification procedure detailed in our previous work~\cite{Kwon2021FirstResults, Kim2023NearQuantum}. 
First, with the Josephson Parametric Amplifier (JPA) unpumped (OFF state), the total JPA-off system noise temperature, $T_{\mathrm{off}}$,
referred to the calibrated noise-source reference plane and including the input thermal noise, was measured via the Y-factor method by sweeping a calibrated noise source from $70\,\text{mK}$ to $150\,\text{mK}$. Next, with the JPA operational at a nominal gain $G_J \approx 21\,\text{dB}$ (ON state), $T_{\text{sys}}$ referred to the cavity output was extracted via the Spectrum Comparison Method (SCM). As derived in Ref.~\cite{Kwon2021FirstResults}, this relationship is expressed as:
\begin{equation}
T_{\text{sys}} = \frac{r \cdot T_{\text{off}}}{\eta_c G_J}
\end{equation}
where $\eta_c$ is the cavity-to-JPA physical transmission efficiency calibrated via \textit{in-situ} VNA reflection measurements, and $r \equiv P_{\text{on}} / P_{\text{off}}$ is the ratio of output power spectral densities between the JPA-ON and JPA-OFF states.

The extracted total system noise temperature ($T_{\text{sys}} \approx 175\text{--}185\,\text{mK}$) reflects the sum of physical noise contributions referred to the cavity output:
\begin{equation}
T_{\text{sys}} = T_{\text{cav}}^{\text{eff}} + T_{\text{JPA}}^{\text{added}} + \frac{T_{\text{post-amp}}}{\eta_c G_J}
\end{equation}
where $T_{\text{cav}}^{\text{eff}} = \frac{h\nu}{k_B} \left( \frac{1}{\exp(h\nu/k_B T_{\text{cav}}) - 1} + \frac{1}{2} \right) \approx 65\,\text{mK}$ is the effective thermal-plus-quantum noise of the cavity at $T_{\text{cav}} = 45\,\text{mK}$ and $\nu = 2.30\,\text{GHz}$ (incorporating $55\,\text{mK}$ zero-point fluctuations). With the intrinsic JPA added noise $T_{\text{JPA}}^{\text{added}} \approx 100\,\text{mK}$ and post-amplifier contributions suppressed below $15\,\text{mK}$ by the high JPA gain, this breakdown confirms that the haloscope operates stably near the quantum-limited sensitivity threshold across the entire scanning band.

\subsection*{Scanning Speed Formulation}
For high-$Q$ cavities where the loaded-cavity bandwidth ($\Delta\nu_c \equiv \nu_c/Q_L$) becomes comparable to the axion signal bandwidth ($\Delta\nu_a$), the standard radiometer equation must account for both line shapes. The detected spectral weighting contains the product of the axion and loaded-cavity Lorentzian profiles, not their convolution. Near resonance, matching the quadratic term of this product gives the following effective-Lorentzian representation:
\begin{equation}
L_a(\nu) \times L_c(\nu) \propto \frac{1}{1 + \left(\frac{\nu - \nu_c}{\nu_c/2Q_a}\right)^2} \frac{1}{1 + \left(\frac{\nu - \nu_c}{\nu_c/2Q_L}\right)^2} \approx \frac{1}{1 + \left(\frac{\nu - \nu_c}{\nu_c/2\sqrt{Q_a^2 + Q_L^2}}\right)^2}
\end{equation}
Using this near-resonance effective-width convention, the corrected detection bandwidth ($\Delta\nu_d$) and frequency step ($\Delta\nu$) are given by:
\begin{equation}
\Delta\nu_d = \frac{\nu_c}{\sqrt{Q_L^2 + Q_a^2}} = \frac{\nu_c}{Q_a} \sqrt{\frac{(1 + \beta)^2}{r_Q^2 + (1 + \beta)^2}}
\end{equation}
\begin{equation}
\Delta\nu = \nu_c \left(\frac{1}{Q_L} + \frac{1}{Q_a}\right) = \frac{\nu_c}{Q_a} \frac{1 + r_Q + \beta}{r_Q}
\end{equation}
where $r_Q \equiv Q_0/Q_a$, $Q_L = Q_0/(1+\beta)$, and $\beta$ is the antenna coupling coefficient. The symbol $\alpha_{\mathrm{em}}$ below denotes the fine-structure constant. Substitution into the adopted scan-rate convention gives the integration time per step:
\begin{equation}
\begin{split}
\Delta t &= \left( g_\gamma^2 \frac{\alpha_{\mathrm{em}}^2}{\pi^2} \frac{\hbar^3 c^3 \rho_a}{\mu_0 \chi} \right)^{-2} \times \left( \frac{1}{\text{SNR}^2} \frac{\omega_c^2 B_{\text{RMS}}^4 V^2 C_{010}^2 Q_a^2}{k_B^2 T_{\text{sys}}^2} \frac{Q_a}{\nu_c} \right)^{-1} \\
&\quad \times \left( \frac{\beta}{1 + \beta} \right)^{-2} \left( \frac{r_Q^2 + (1 + \beta)^2}{(1 + \beta)^2} \right)^{-1/2} \left( \frac{r_Q}{1 + r_Q + \beta} \right)^{-2}
\end{split}
\end{equation}
where $g_\gamma$ is the dimensionless, model-dependent axion-photon coupling coefficient. We use $\rho_a \approx 0.45\,\text{GeV/cm}^3$ for the local dark-matter density, and $C_{010}$ is the $TM_{010}$ form factor. The fully corrected scanning speed is thus formulated as:
\begin{equation}
\begin{split}
\frac{d\nu}{dt} &= \left( g_\gamma^2 \frac{\alpha_{\mathrm{em}}^2}{\pi^2} \frac{\hbar^3 c^3 \rho_a}{\mu_0 \chi} \right)^2 \times \left( \frac{1}{\text{SNR}^2} \frac{\omega_c^2 B_{\text{RMS}}^4 V^2 C_{010}^2 Q_a^2}{k_B^2 T_{\text{sys}}^2} \right) \\
&\quad \times \left( \frac{\beta}{1 + \beta} \right)^2 \left( \frac{r_Q^2 + (1 + \beta)^2}{(1 + \beta)^2} \right)^{1/2} \frac{r_Q}{1 + r_Q + \beta}
\end{split}
\end{equation}
Writing $x \equiv Q_L/Q_a = r_Q/(1+\beta)$, the finite-linewidth factor in the last line simplifies to $\mathcal{F}(x)=\frac{\sqrt{1+x^2}x}{1+x}$. By factoring out the linear $x$ dependence (which corresponds to the standard $Q_0$ scaling), we can express $\mathcal{F}(x) = x \cdot \mathcal{F}_{\text{alt}}(x)$, where $\mathcal{F}_{\text{alt}}(x) \equiv \frac{\sqrt{1+x^2}}{1+x}$ explicitly isolates the pure finite-linewidth correction factor utilized in Table~4. Notably, $\mathcal{F}(x)$ exhibits a linear dependence, $\mathcal{F}(x) \approx x$, in both extreme limits ($x \ll 1$ and $x \gg 1$). In the conventional low-$Q_L$ regime ($x \ll 1$), substituting $x = Q_0/[Q_a(1+\beta)]$ recovers the standard radiometer scan rate $\frac{d\nu}{dt} \propto \frac{B_{\text{RMS}}^4 V^2 C_{010}^2 Q_0}{T_{\text{sys}}^2} \frac{\beta^2}{(1+\beta)^3}$, reconciling with the individual parameter contribution factor breakdown in Table~\ref{tab:comparison}. Crucially, in the ultra-high-$Q_L$ regime ($x \gg 1$), $\mathcal{F}(x)$ remains strictly linear in $Q_L$, confirming that cavity quality factor enhancements continue to yield an unsuppressed linear scan-rate advantage even when the resonator bandwidth is far narrower than the dark matter linewidth, as independently derived by Cervantes et al.~\cite{Cervantes2024PRD}. Through the intermediate crossover region ($x \sim 1$), the full analytical expression smoothly interpolates between these asymptotic limits. Here, $T_{\text{sys}}$ is the experimentally measured, receiver-referred effective system noise temperature at the operating coupling. Accordingly, the separate coupling-dependent thermal-noise transfer factor used when decomposing physical and added noise contributions~\cite{Kim2020AxionScanRate} is not applied again in this convention, ensuring no double-counting of signal attenuation.

\subsection*{Data Analysis}

The spectral data processing and candidate selection pipeline followed the established CAPP analysis framework~\cite{Kim2023NearQuantum, ahn2024prx, Kwon2021FirstResults}. Initially, the power spectrum baseline, which accounts for the effects of frequency-dependent gain and impedance mismatch within the receiver chain, was removed. To mitigate signal attenuation during this baseline removal, we employed a 5th-order Savitzky-Golay filter applied in a dual-pass strategy. Extensive Monte Carlo simulations utilizing the simulated axion line shape ($Q_a \approx 10^6$, $\Delta\nu_a \approx 2.3\,\text{kHz}$) systematically optimized the dual window sizes to 181 and 161 bins. While a conventional single-pass filter resulted in a signal attenuation penalty exceeding 60\%, this optimized dual-filter strategy successfully reduced the penalty to below 50\%, yielding an effective signal-to-noise retention efficiency of $\eta_{\text{BL}} \approx 52\%$.

Subsequently, the baseline-corrected spectra were combined to construct a grand spectrum across the entire scan range by convolving them with the noise-normalized axion line shape. For this convolution, the axion velocity dispersion relative to the Galactic Center was set to $270\,\text{km/s}$, and the Sun's orbital speed around the Galaxy was set to $230\,\text{km/s}$, following the Standard Isothermal Sphere Halo Model~\cite{Turner1990AxionLine}.

Next, axion candidates were identified and examined using a threshold-based criterion. For the combined data from all three runs, a target signal-to-noise ratio (SNR) of $5\sigma$ with a 90\% local detection confidence level statistically translates to an operational candidate selection threshold of $3.718\sigma$. Throughout the runs, five spectral peaks exceeded this threshold. Each candidate was subjected to a dedicated 10-hour re-scan procedure; none persisted with the expected dark matter signal behavior, and all were therefore safely excluded as axion candidates.

Finally, exclusion limits on the axion-photon coupling constant ($\vert{}g_{a\gamma\gamma}\vert{}$) were derived. The total systematic uncertainty budget on the expected signal power was evaluated from statistically independent contributions: magnetic field profile mapping (2\%), cavity quality factor measurements (5\%), noise temperature calibration (3\%), and form factor electromagnetic simulations (3\%). Because the axion conversion power scales quadratically with the magnetic field ($P_a \propto B^2$), the fractional field uncertainty enters the power normalization at twice its value. Thus, the overall systematic uncertainty ($\sigma_{\mathrm{sys}}$) was rigorously propagated via quadrature sum:
\begin{equation}
\sigma_{\mathrm{sys}} = \sqrt{\left(2 \times 2\%\right)^2 + \left(5\%\right)^2 + \left(3\%\right)^2 + \left(3\%\right)^2} = \sqrt{59}\% \approx 7.68\% \simeq 8\%
\end{equation}
This nominal 8\% systematic allowance was directly incorporated into the final $90\%$ confidence exclusion limit derivation.

\section*{Data Availability}
The data supporting the findings of this study, including the data underlying the figures and axion exclusion limits, are available
from the corresponding authors upon reasonable request.

\section*{Code Availability}
The custom Python routines used for the nonlinear Gittleman-Rosenblum model fitting, the dual-pass Savitzky-Golay baseline removal, and the final limit construction to generate the main findings of this study are available from the corresponding authors upon reasonable request. 

\section*{Author Contributions}
D.A., O.K., and W.C. conceived and designed the experiments and executed the CAPP-PACE haloscope physics runs. D.A., O.K., S.P., and D.Y. constructed the HTS cavities. D.A., J.K., J.L., and H.B. performed cryogenic microwave characterizations. S.U., A.F.v.L., and Y.N. contributed to quantum-noise-limited JPA production and operation. D.A. and J.K. analyzed the data and developed the analytical scan-rate models. Y.N., S.Y., and Y.K.S. contributed to data interpretation and manuscript review. D.A., O.K., and W.C. wrote the manuscript with contributions and feedback from all co-authors. O.K., W.C., and Y.K.S. supervised the project.

\section*{Competing Interests}
The authors declare no competing interests.

\section*{Acknowledgement}
This work was generously supported by the Institute for Basic Science. Specifically, the experimental work and data collection were funded by Grant No. IBS-R017-D1, while manuscript preparation and publication were supported by Grant No. IBS-R040-C1-2025-a00. Additional support was provided by JSPS KAKENHI (Grant No. JP22H04937). A.F.v.L. acknowledges the support of a JSPS postdoctoral fellowship.

During the preparation of this work, the authors used Gemini for language polishing and text refinement. The authors reviewed and edited the content as needed and take full responsibility for the publication's contents.

\newpage

\bibliographystyle{naturemag}
\bibliography{reference}

\newpage

\section*{Tables}

\begin{table}[htbp]
\centering
\resizebox{\textwidth}{!}{
\begin{tabular}{@{}clccccc@{}}
\toprule
\textbf{Tape no.} & \textbf{Stabilizer} & \textbf{Protection layer} & \textbf{HTS layer} & \textbf{Buffer layer} & \textbf{Substrate} & \textbf{Manufacturer} \\ \midrule
1 & Cu ($100\,\mu\text{m}$) & Ag ($1.5\,\mu\text{m}$) & GdBCO ($3{\sim}5\,\mu\text{m}$) & MgO ($3.5\,\mu\text{m}$) & Hastelloy ($50\,\mu\text{m}$) & THEVA \\
2 & Cu ($20\,\mu\text{m}$) & Ag ($2\,\mu\text{m}$) & EuBCO + BHO ($2.5\,\mu\text{m}$) & MgO ($0.7\,\mu\text{m}$) & Hastelloy ($50\,\mu\text{m}$) & Fujikura \\ \bottomrule
\end{tabular}
}
\caption{\textbf{Specifications of commercial biaxially-textured REBCO coated conductors.} Layer architecture and nominal thicknesses of the HTS tapes utilized in this study. Selective mechanical delamination removes the $50\,\mu\text{m}$ Hastelloy substrate, utilizing the copper stabilizer as a low-loss conductive backing. The EuBCO tape incorporates $\text{BaHfO}_3$ (BHO) nanorods acting as artificial pinning centers (APCs).}
\label{tab:tapes}
\end{table}

\newpage

\begin{table}[htbp]
\centering
\resizebox{\textwidth}{!}{
\begin{tabular}{@{}ccllccccc@{}}
\toprule
\textbf{No.} & \textbf{Architecture} & \textbf{HTS Film} & \textbf{Alignment} & \textbf{$f_0$ (GHz)} & \textbf{$D \times H$ (mm)} & \textbf{$V$ (L)} & \textbf{$n_{\text{gaps}}$} & \textbf{Saturated $Q_0$ (8 T)} \\ \midrule
1 & Aluminum (Arch A) & YBCO ($0.8\,\mu\text{m}$) & Precision Wedge & 6.93 & $33 \times 100$ & 0.08 & 12 & $3.3 \times 10^5$ \\
2 & Brass/Cu (Arch B) & GdBCO ($3\text{--}5\,\mu\text{m}$) & Inner Mold & 2.30 & $100 \times 190$ & 1.50 & 32 & $5.0 \times 10^5$ \\
3 & Aluminum (Arch A) & EuBCO + BHO ($2.5\,\mu\text{m}$) & Precision Wedge & 2.27 & $100 \times 190$ & 1.50 & 34 & $3.5 \times 10^6$ \\
4 & Aluminum (Arch A) & EuBCO + BHO ($2.5\,\mu\text{m}$) & Precision Wedge & 5.40 & $42 \times 90$ & 0.12 & 14 & $1.4 \times 10^7$ \\ \bottomrule
\end{tabular}
}
\caption{\textbf{Geometric and operational parameters of fabricated HTS cavities.} Summary of 3D polygonal resonant cavities operating in the $TM_{010}$ mode. $D$ and $H$ denote inner diameter and active length, $V$ represents effective cavity volume, $n_{\text{gaps}}$ is the number of longitudinal assembly seams, and $f_0$ is the un-tuned resonant frequency. Architecture A employs precision aluminum wedges (Nos. 1, 3, 4), while Architecture B utilizes an inner-mold frame optimized for volume and mK thermalization (No. 2). Note that Cavity No.~1 represents our first-generation prototype~\cite{ahn2022biaxially} using as-received, non-delaminated YBCO tapes with Ni-9W substrates, whereas Cavities Nos.~2--4 incorporate conductive copper backings produced via mechanical delamination. For No.~2, the tabulated $5.0 \times 10^5$ value represents the empty-cavity measurement without the sapphire tuning rod, while the operational axion-run configuration incorporating the tuning rod yielded a representative $Q_0 \simeq 4.1 \times 10^5$.}
\label{tab:cavities}
\end{table}

\newpage

\begin{table}[htbp]
\centering
\begin{tabular}{@{}lccc@{}}
\toprule
\textbf{Parameter} & \textbf{Run 1 (2021)} & \textbf{Run 2 (2021)} & \textbf{Run 3 (2022)} \\ \midrule
Period & Oct 01 - Oct 16 & Oct 26 - Nov 4 & Jul 20 - Aug 24 \\
Frequency range (GHz) & 2.284 - 2.295 & 2.283 - 2.294 & 2.282 - 2.293 \\
Frequency bin width & \multicolumn{3}{c}{100 Hz} \\
Single tuning step & \multicolumn{3}{c}{4 kHz} \\
Sweep time & \multicolumn{3}{c}{1 sec} \\
Number of spectra & 180 & 220 & 520 \\
Unloaded quality factor ($Q_0$) & \multicolumn{3}{c}{$385,000$--$427,000$ ($Q_{0,\text{RMS}}$: $4.05 \times 10^5$)} \\
Loaded quality factor ($Q_L$) & \multicolumn{3}{c}{$147,000$--$155,000$ (measured \textit{in-situ})} \\
Average Magnetic field ($B_{\text{RMS}}$) & \multicolumn{3}{c}{6.95 T (at $B_{\text{max}} = 8.0\,\text{T}$)} \\
Cavity volume ($V$) & \multicolumn{3}{c}{1.5 L} \\
Form factor ($C_{010}$) & \multicolumn{3}{c}{0.58 - 0.62} \\
MXC temperature & \multicolumn{3}{c}{21 mK} \\
Cavity temperature & \multicolumn{3}{c}{43--45 mK} \\
System noise ($T_{\text{sys}}$) & 175 mK & 185 mK & 175 mK \\ \bottomrule
\end{tabular}
\caption{\textbf{Operational history and cryogenic noise performance of the CAPP-PACE haloscope runs.} Summary of physics data acquisition runs using the 2.3 GHz GdBCO pathfinder cavity (Architecture B) in a 6.95 T volume-averaged magnetic field ($B_{\text{max}} = 8.0\,\text{T}$). Both $Q_L$ and coupling $\beta$ were measured \textit{in-situ} in real-time at every tuning step during live physics data acquisition. Physical cavity temperatures of $43{-}45\,\text{mK}$ enabled a stable system noise temperature ($T_{\text{sys}}$) of $175{-}185\,\text{mK}$ across all runs. The data acquisition encompasses a cumulative live integration time of approximately 30 days across the physics runs, after excluding cryogenic cycling, maintenance, and tuning overheads.}
\label{tab:runs}
\end{table}

\newpage

\begin{table}[htbp]
\centering
\resizebox{\textwidth}{!}{
\begin{tabular}{@{}lccc@{}}
\toprule
\textbf{Parameter / Metric} & \textbf{2020 Copper Haloscope~\cite{Kim2023NearQuantum}} & \textbf{HTS Haloscope (This Work)} & \textbf{Scan-Rate Contribution Factor} \\ \midrule
Representative Frequency ($\nu$) & $2.2852\,\text{GHz}$ & $2.2885\,\text{GHz}$ & --- \\
Volume-Averaged Field ($B_{\text{RMS}}$) & $7.20\,\text{T}$ & $6.95\,\text{T}$ & \textbf{0.868} \quad ($(B_{\text{HTS}}/B_{\text{Cu}})^4$) \\
Active Cavity Volume ($V$) & $1.12\,\text{L}$ & $1.50\,\text{L}$ & \textbf{1.794} \quad ($(V_{\text{HTS}}/V_{\text{Cu}})^2$) \\
Form Factor ($C_{010}$) & $0.45$ & $0.60$ & \textbf{1.778} \quad ($(C_{\text{HTS}}/C_{\text{Cu}})^2$) \\
Quality Factor ($Q_0/Q_L$) & $8.8\times 10^4/3.0 \times 10^4$ & $4.05\times 10^5/1.50 \times 10^5$ & \textbf{4.602} \quad ($Q_{0,\text{HTS}}/Q_{0,\text{Cu}}$) \\
Antenna Coupling Factor ($\frac{\beta^2}{(1+\beta)^3}$) & $0.1481$ ($\beta=1.93$) & $0.1468$ ($\beta=1.70$) & \textbf{0.991} \quad ($\text{Coupling}_{\text{HTS}}/\text{Coupling}_{\text{Cu}}$) \\
Effective System Noise ($T_{\text{sys}}$) & $234\,\text{mK}$ & $177\,\text{mK}$ & \textbf{1.748} \quad ($(T_{\text{sys,Cu}}/T_{\text{sys,HTS}})^2$) \\ \midrule
\textbf{Uncorrected Raw Hardware Speedup} & --- & --- & \textbf{22.1-fold} \quad ($\prod$ Raw hardware factors) \\
Finite Linewidth Penalty ($\mathcal{F}_{\text{alt}}$) & $0.971$ & $0.879$ & \textbf{0.905} \quad ($\text{Linewidth}_{\text{HTS}}/\text{Linewidth}_{\text{Cu}}$) \\ \midrule
\textbf{Ideal Hardware Scan-Rate Speedup} & --- & --- & \textbf{20.0-fold} \quad ($22.1 \times 0.905$) \\
Baseline Signal Retention Penalty ($\eta_{\text{BL}}^2$) & $0.640$ ($\eta_{\text{BL}}=0.80$) & $0.267$ ($\eta_{\text{BL}}=0.517$) & \textbf{0.417} \quad ($(\eta_{\text{BL,HTS}}/\eta_{\text{BL,Cu}})^2$) \\ \midrule
\textbf{Analysis-Adjusted Scan-Rate enhancement} & --- & --- & \textbf{8.3-fold} \quad ($20.0 \times 0.417$) \\ \bottomrule
\end{tabular}
}
\caption{\textbf{Quantitative parameter and scan-rate enhancement breakdown between conventional copper and HTS haloscopes.} Detailed 1:1 comparison of operational parameters alongside their exact physical scan-rate contribution factors between the 2020 CAPP-PACE copper baseline~\cite{Kim2023NearQuantum} and the 2021-2022 HTS pathfinder haloscope. The raw hardware contribution product yields a 22.1-fold speedup under the low-$Q_L$ approximation using $Q_0$ and coupling factor $\frac{\beta^2}{(1+\beta)^3}$. Incorporating the finite axion linewidth penalty factor $\mathcal{F}_{\text{alt}}(x)$ for $Q_a=10^6$ ($x \equiv Q_L/Q_a$, see Methods; yielding a relative contribution factor of $0.905$) refines the ideal hardware speedup to 20.0-fold. Applying the baseline signal retention penalty ($0.417$) ultimately yields an 8.3-fold analysis-adjusted scan-rate advantage.}
\label{tab:comparison}
\end{table}
\newpage

\section*{Figure Legends}

\newpage

\newpage

\newpage

\newpage

\clearpage


\singlespacing
\nolinenumbers
\makeatletter
\let\title\combinedsavedtitle
\let\author\combinedsavedauthor
\let\date\combinedsaveddate
\gdef\AB@authlist{}
\gdef\AB@affillist{}
\gdef\AB@authors{}
\gdef\AB@author{}
\gdef\AB@las{}
\gdef\AB@lasx{\protect\Authand}
\gdef\AB@as{}
\gdef\AB@au@str{}
\setcounter{authors}{0}
\setcounter{affil}{0}
\newaffiltrue
\makeatother

\title{\textbf{Supplementary Information for:\\ Multi-tesla operation of high-temperature superconducting cavities for accelerated axion dark matter searches}}

\makeatletter
{{\renewenvironment{tabular}[2][]{\begin{center}}{\end{center}}\AB@maketitle}}
\makeatother

\newpage

\section*{Supplementary Note: High-Field $Q_0 = 1.4 \times 10^7$ Joint Resonance Fit}

To rigorously confirm the saturated unloaded quality factor $Q_0 \approx 1.4 \times 10^7$ at $8\,\text{T}$ for the $5.40\,\text{GHz}$ EuBCO cavity (No.~4 in Table~\ref{tab:cavities}), a joint scalar resonance fit was performed on the measured $|S_{21}|$, $|S_{11}|$, and $|S_{22}|$ logarithmic magnitudes. The data were acquired across 1,001 frequency points with an intermediate frequency bandwidth (IFBW) of $1\,\text{kHz}$, and analyzed using a Fano asymmetric line-shape model~\cite{Fano1961, Khalil2012}.

For each transmission or reflection channel $k \in \{21, 11, 22\}$, the fitted scalar response function is given by:
\begin{equation}
y_k(f) = C_k + m_k u + 10 \log_{10} \left[ \frac{(\epsilon+q_k)^2 + r_k^2}{1+\epsilon^2} \right]
\end{equation}
where $\epsilon = 2(f-f_0)/\Gamma$ is the normalized detuning, $u$ is the normalized frequency coordinate across the sweep span, $C_k$ and $m_k u$ account for the baseline offset and linear background tilt, while $q_k$ and $r_k$ characterize the Fano resonance asymmetry. During the simultaneous non-linear least-squares optimization, each channel was weighted by a residual scale estimated from its median absolute deviation.

The joint fit yielded a loaded pole linewidth of $\Gamma \approx 401\,\text{Hz}$, corresponding to $Q_L \approx 1.35 \times 10^7$. Accounting for the weak total external coupling ($\beta_{\text{tot}} = \beta_1 + \beta_2 \approx 0.058$), the derived unloaded quality factor is $Q_0 \approx 1.42 \times 10^7$. The uncertainties of these extracted parameters are dominated by systematic measurement and calibration errors, conservatively bounded by the $\sim 5\%$ margin applied in the main text analysis.

\end{document}